\documentclass[12pt,preprint]{aastex}

\newcommand{\hi}{\ion{H}{1}}			
\newcommand{\hii}{\ion{H}{2}}			
\newcommand{\hst}{{\it HST}}			
\newcommand{\kms}{km~s$^{-1}$}			
\newcommand{\msun}{\mbox{${\cal M}_{\odot}$}}	
\newcommand{\zo}{Z{$_\odot$}}			
\newcommand{\rprime}{\mbox{$r^\prime$}}         
\newcommand{\gprime}{\mbox{$g^\prime$}}         
\newcommand{\iprime}{\mbox{$i^\prime$}}         
\newcommand{\tlabel}{$\times$ 1e14~erg~cm$^{-2}$~s$^{-2}$~\AA$^{-1}$}
\newcommand{\Oiii}{[OIII]$_{\lambda4959+5007}$}
\newcommand{\Oiiiu}{[OIII]$_{\lambda5007}$}
\newcommand{\Nii}{[NII]$_{\lambda6548+6584}$}
\newcommand{\Sii}{[SII]$_{\lambda6713+6731}$}

\shorttitle{Stellar Clusters in Stephan's Quintet}
\shortauthors{Trancho et al.}

\begin{document}

\title{Gemini Spectroscopic Survey of Young Star Clusters in Merging/Interacting Galaxies. IV. Stephan's Quintet}  


\author{Gelys Trancho}
\affil{Gemini Observatory, Casilla 603, La Serena, Chile}
\and
\affil{Giant Magellan Telescope Organization, 251S. Lake ave., Pasadena 91001,USA}
\email{gtrancho@gmto.org}
\author{Iraklis S. Konstantopoulos}
\affil{Department of Astronomy \& Astrophysics, 525 Davey Lab., The Pennsylvania State University, University~Park, PA 16802, USA}
\author{Nate Bastian}
\affil{Excellence Cluster Universe, Boltzmannstr. 2, 85748 Garching, Germany}
\affil{School of Physics, University of Exeter, Stocker Road, Exeter EX4 4QL, United Kingdom}
\author{Konstantin Fedotov \& Sarah Gallagher}
\affil{Physics and Astronomy Department, University of Western Ontario, London, ON N6A 3K7, Canada}
\author{Brendan Mullan \& Jane C. Charlton}
\affil{Department of Astronomy \& Astrophysics, 525 Davey Lab., The Pennsylvania State University, University~Park, PA 16802, USA}

\clearpage
\begin{abstract}

We present a spectroscopic survey of 21 young massive clusters and complexes and one tidal dwarf galaxy candidate (TDG) in Stephan's Quintet, an interacting compact group of galaxies. All of the selected targets lie outside the main galaxies of the system and are associated with tidal debris.  We find clusters with ages between a few and 125~Myr and confirm the ages estimated through \hst\ photometry by Fedotov et al.~(2011), as well as their modelled interaction history of the Quintet. Many of the clusters are found to be relatively long-lived, given their spectrosopically derived ages, while their high masses suggest that they will likely evolve to eventually become intergalactic clusters. One cluster, T118, is particularly interesting, given its age ($\sim125$~Myr), high mass ($\sim2 \times10^6$~M$_{\odot}$) and position in the extreme outer end of the young tidal tail.  This cluster appears to be quite extended (R$_{\rm eff} \sim 12-15$~pc) compared to clusters observed in galaxy disks (R$_{\rm eff} \sim 3-4$~pc), which confirms { an effect} we previously found in the tidal tails of NGC~3256, where clusters are similarly extended.  We find that star and cluster formation can proceed at a continuous pace for at least $\sim150~$Myr within the tidal debris of interacting galaxies.  The spectrum of the TDG candidate is dominated by a young population ($\sim7$~Myr), and assuming a single age for the entire region, has a mass of at least~$10^6$~M$_{\odot}$. 

\end{abstract}

\keywords{galaxies: star clusters --- galaxies: groups: Stephan's Quintet -- galaxies: interactions}

\section{INTRODUCTION}

The goal of the {\it Gemini Spectroscopic Survey of Young  Star Clusters in Merging/Interacting Galaxies} (Trancho et al.~2007a) is to use clusters as tracers of the interaction history and therefore the evolution of their host galaxies.  Using optical spectroscopy we have derived the ages and metallicities of dozens of massive clusters in nearby ongoing mergers.  While spectroscopy is much more expensive than photometry in terms of telescope time, it allows for precise,  non-degenerate age/metallicity derivations and also offers the advantage of kinematical information that can be used to identify sub-populations within the merger/interaction.  

Thus far, we have focused on ongoing major mergers between spiral galaxies, namely NGC~3256 (Trancho et al.~2007a,b; hereafter T07a,b) and the Antenn\ae\ galaxies (NGC~4038/39; Bastian et al.~2009; hereafter B09).  In NGC~3256 we identified and characterised a number of young, massive clusters in the tidal tails and the main body of the galaxy.  The clusters in the tidal debris have ages between $\sim80$ and $\sim200$~Myr and with effective (half-light) radii of R$_{\rm eff} = 10-20$~pc are much more extended than young clusters in the parent galaxy disks (typical sizes of $\sim4-9~$pc). The clusters in the main body of the galaxy are kinematically associated with the disk (i.\,e. rotating along with it), despite the advanced state of the merger. 

Our work on the Antenn\ae~(B09) found evidence of an increasing star formation history over the past few hundred Myr, in accord with numerical models of the galaxy merger~(Mihos et al.~1993; Karl et al.~2010; although see Karl et al.~2011 for a contrary view). Additionally, we found clusters in the projected outer disks of the interacting galaxies have ``halo" type kinematics, showing the randomization of stellar orbits. These clusters are likely to contribute to the buildup of a stellar halo around the future merger remnant. Simulations of cluster formation in galaxy mergers have shown that the majority of cluster formation during the merger occurs in disks (either in the progenitor disks or a disk formed due to the infall of gas in the center of the remnant; Kruijssen et al.~2011), in agreement with the observations presented in this series (T017a,b; B09).  In the simulations, however, clusters that form in the progenitor disks that are subsequently moved into the halo, which agrees with observations of clusters in advanced stage mergers (e.g., Schweizer \& Seitzer~1998; Schweizer et al.~2004).

In the current study, we shift our focus to the more complicated environment of a group of interacting galaxies, Stephan's Quintet { (hereafter SQ). SQ, a compact group, has been the object of a number of studies, which have treated it as whole (e.\,g. \citealt{Moles1997}; \citealt{Sulentic2001}; \citealt{Gallagher2001}; \citealt{Williams2002}), or have focussed on certain aspects. Primary among those are its tidal features and star formation in the intra-group medium (e.\,g. \citealt{Xu1999}; \citealt{Mendes2001}; \citealt{Lisenfeld2004}; \citealt{Xu2005}; \citealt{Torres2009}) and the $\sim$ 40 kpc long X-ray-emitting shock front, the result of a high-speed collision between two of its members (e.\,g. \citealt{Xu2003}; \citealt{Trinchieri2003}; \citealt{Appleton2006}; \citealt{Guillard2010}; \citealt{Suzuki2011}).

In this paper, we concentrate on the clusters that are outside the nominal host galaxies, i.\,e. part of the tidal debris of the system. To avoid the type of ionization and emission-line broadening that comes with shocks, most of the objects we study lie in areas away from the mentioned X-ray shock front -- although four slits lie in regions that may be affected by the underlying shock-emission (see Section~\ref{sec:e-spec}). Previous studies (see above) have revealed a large number of young intergalactic clusters in SQ. }
 
Some of the clusters that have been found outside galaxies by other works appear to have been formed in the tidal debris released by interactions (e.g. Gallagher et al.~2001; Tran et al.~2003; Bastian et al.~2005; Trancho et al.~2007a; Werk et al.~2008), while others appear to have formed in the outer disks of the progenitor spirals and have been thrown into the concurrently forming stellar halo (Schweizer et al.~2004; Bastian et al.~2009). What fraction of these clusters end up in the stellar halo of the resultant merged pair or group, and what fraction are ejected into intra-group space is currently an open question.

Our study is designed to complement a Hubble Space Telescope~(\hst) Wide-Field Camera~3 (WFC3) photometric study presented by Fedotov et al.~(2011, hereafter F11) that reaches fainter clusters and thus obtains a more global picture of the full cluster population. Additionally, the WFC3 images provide the opportunity to search for fainter clusters near our spectroscopic targets as well as to compare the background (i.\,e.~light from tidal features) colors with the identified targets.  This new paper is, however, crucial for validating the findings of its more far-reaching counterpart, which is based on BVI photometry. Lacking the non-degenerate diagnostic power of U-band (or spectroscopy), the Fedotov~et~al. study requires independently derived age measurements. We will refer to F11 throughout this paper, and we largely confirm their results. In the present paper we study the properties of 21 clusters/complexes and a Tidal Dwarf Galaxy candidate that are found in the tidal debris of the interacting system, SQ.  

The study is organized in the following way; in \S~\ref{sec:obs} we introduce the spectroscopic and photometric observations used in this study.  In \S~\ref{sec:properties} we derive the properties (age, mass, extinction, metallicity, and kinematics) of each of the clusters and in \S~\ref{sec:results} we discuss our results in the context of star/cluster formation in tidal debris and compare our findings with those of tidal dwarf galaxies.  Our conclusions are presented in \S~\ref{sec:conclusions}.

\section{OBSERVATIONS AND REDUCTION}
\label{sec:obs}
\subsection{Gemini-GMOS Spectroscopy}
\label{sec:spectroscopy}
We  obtained spectra of 40 candidate sources in SQ using the MOS mode of GMOS on Gemini North.   Imaging and spectroscopy of star clusters in SQ were obtained with GMOS-N in semesters 2004B and 2006A. The data were obtained as part of Directory Discretionary
program GN-2004B-DD-8 and  the queue program GN-2006A-Q-38.  Imaging was obtained from the Gemini Public Outreach program in three
filters \gprime, \rprime and  \iprime.  Two GMOS masks were used for the spectroscopy. The selection of star cluster candidates was based on colors and magnitudes from the pre-imaging and was supplemented by the candidate list of Gallagher et al. (2001, here after G01) which was based on HST/WFPC2 imaging. Only sources that were bright enough to obtain high enough S/N in the time allocated to obtain reliable velocities and stellar population properties were included.  Additionally, priority was given to candidates that lie (in projection) upon known tidal features, in order to increase the probability of being associated with the Stephan's Quintet system.  All candidates (with the exception of the Tidal Dwarf Galaxy candidate which is discussed in \S~\ref{sec:tdg}) are dominated by a point-like (i.e. unresolved) source on the ground-based images, although based on HST imaging, the youngest clusters are in fact complexes of clusters, similar to what has been seen in other environments (e.g.~Bastian et al.~2005, 2009).

\begin{figure*}
 \begin{center}
 	\epsscale{1.0}
 	\plotone{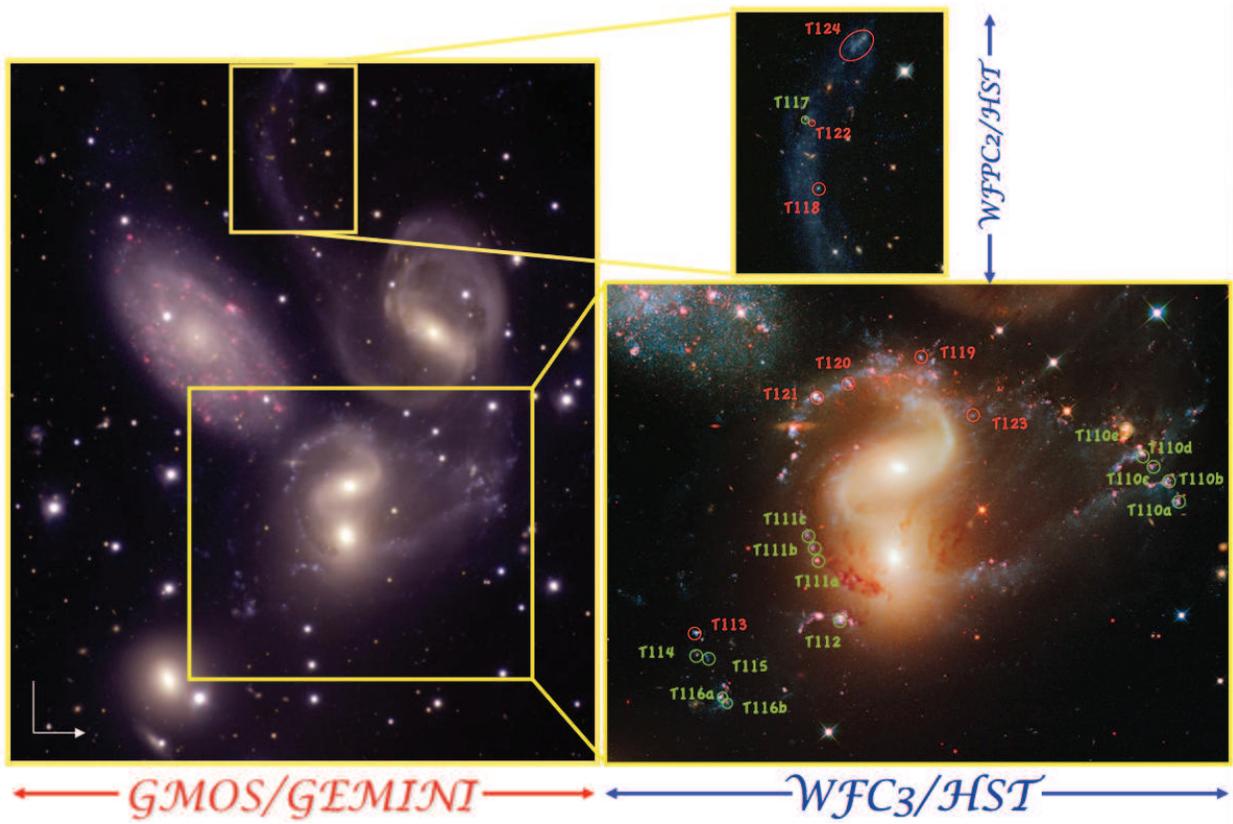}
 	\caption{Gemini/HST images of Stephan's Quintet.  The observed clusters are shown where red represents clusters which display an absorption dominated spectrum, while green represents emission dominated spectra.  A zoomed in image of each cluster is shown in Fig.~\ref{fig:images}.} 
    	 \label{fig:image}
 \end{center} 
 \end{figure*}

\begin{figure*}
 \begin{center}
 	\epsscale{1.0}
 	\plotone{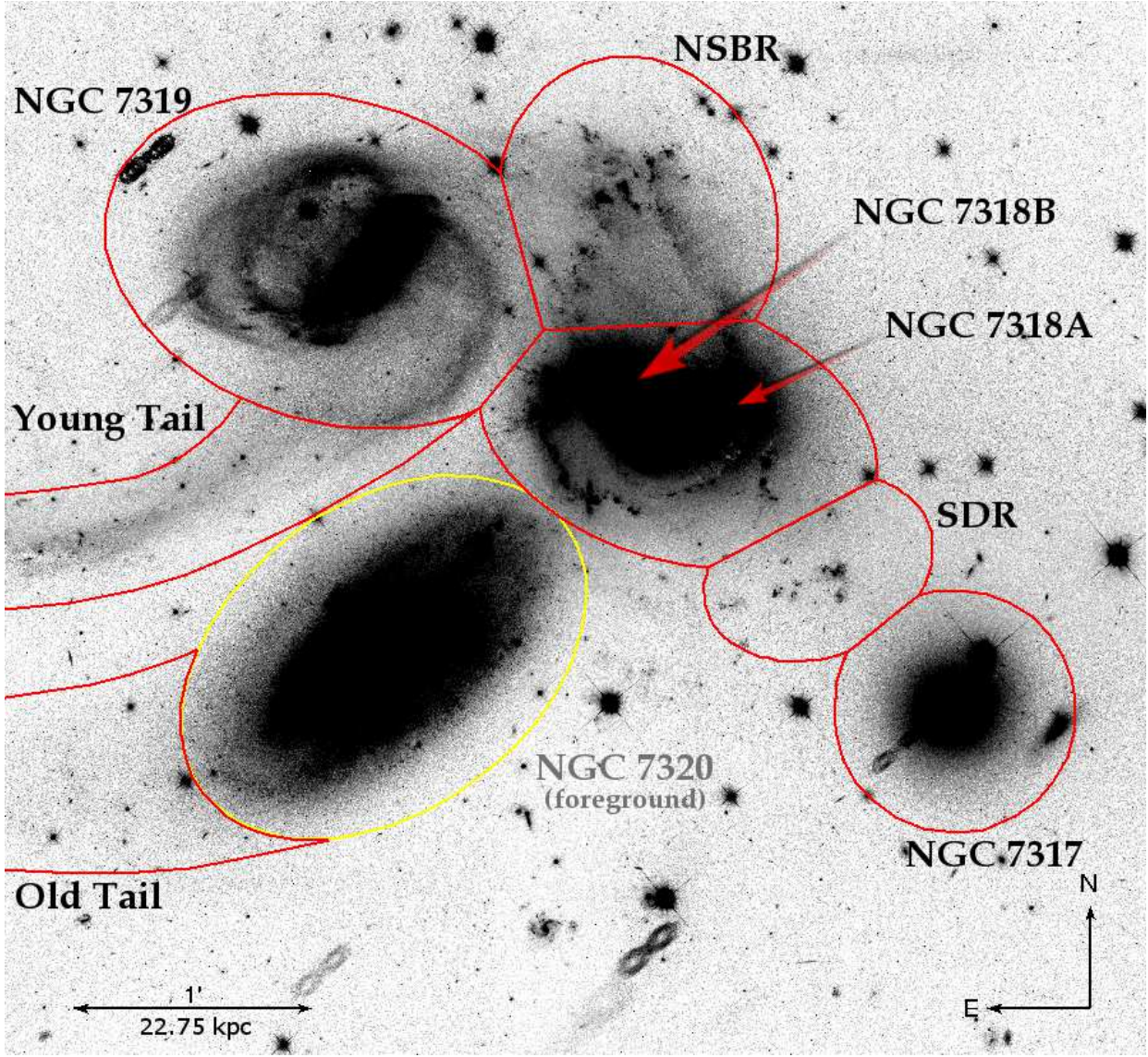}
 	\caption{WFC3 { F606W} image of Stephan's Quintet with the regions studied highlighted.} 
    	 \label{fig:image_label}
 \end{center} 
 \end{figure*}

The B600 grating was used with a slit width of 0.75~arcsec, resulting in an instrumental resolution of 110 km/s at 5000 \AA. The spectroscopic observations were obtained as 8 individual exposures with an exposure time of 1600 sec each. Sky subtraction was performed locally, in apertures within each slit. The extent of these apertures varies according to the width of ``sky" available. This ensures the removal of signatures from diffuse components, which are expected to be present in the complex intra-group medium of SQ. After sky subtraction, the spectra were flux calibrated using exposures of spectrophotometric standard star G191B2B. Of the 40 candidates targeted for spectroscopy, 22 were confirmed to be part of the SQ system based on their radial velocities.  Details of the confirmed sources (IDs, positions, and magnitudes) are given in Table~\ref{table:properties}.  Figure~\ref{fig:image} shows a Gemini/\hst\ image of SQ, with the observed candidate clusters marked with their ID numbers.  Additionally, in Fig.~\ref{fig:image_label} we show a HST/WFC3 image of SQ, and label specific regions that will be studied in detail in \S~\ref{sec:results}. 

\begin{figure}
 \begin{center}
 	\includegraphics[width=12cm,angle=90]{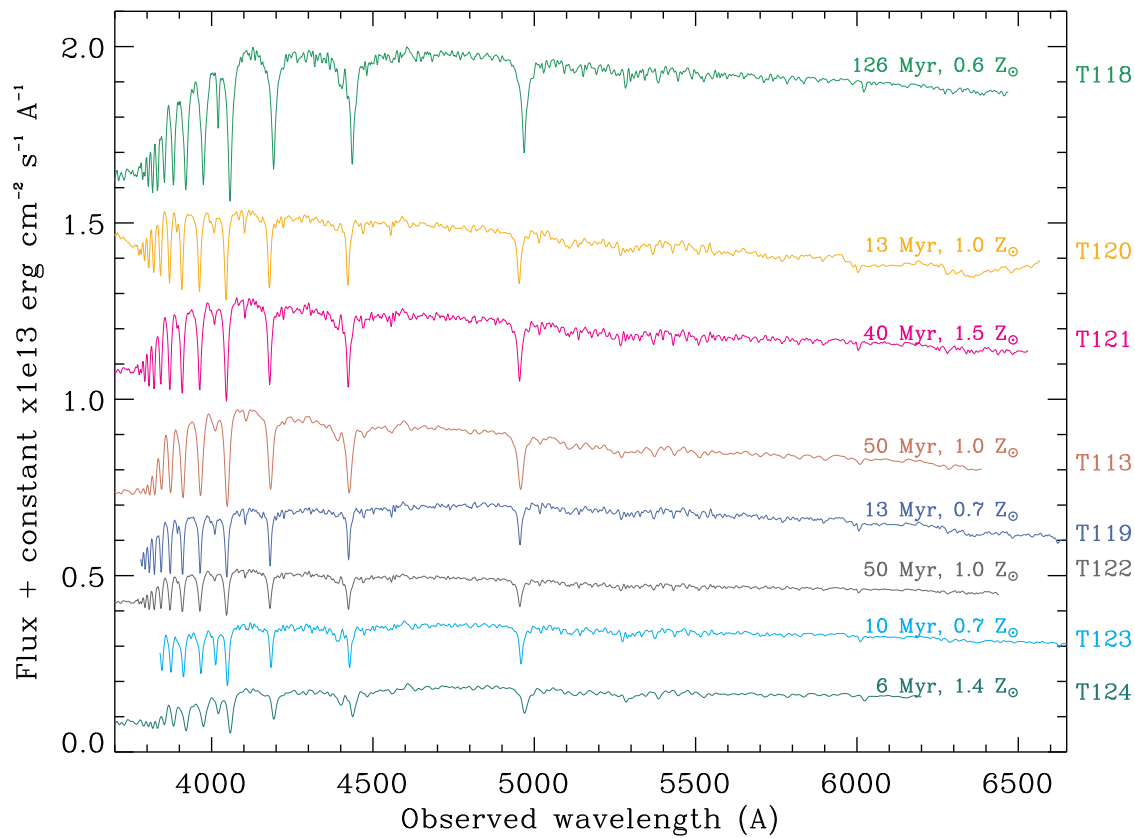}
 	\caption{Absorption line spectra of seven clusters/clumps, with  ages in Myr and metallicities as a fraction of solar labelled. ID T118 (top) is the massive cluster situated in the young tail, discussed in Section~\ref{sec:t118}.  ID T124 (bottom) marks the candidate Tidal Dwarf Galaxy candidate at the tip of the young tail (see Section~\ref{sec:tdg}). 
	}\label{fig:spec1}

 \end{center} 
 \end{figure}

\begin{figure}
 \begin{center}
 	\epsscale{1.0}

 	\plotone{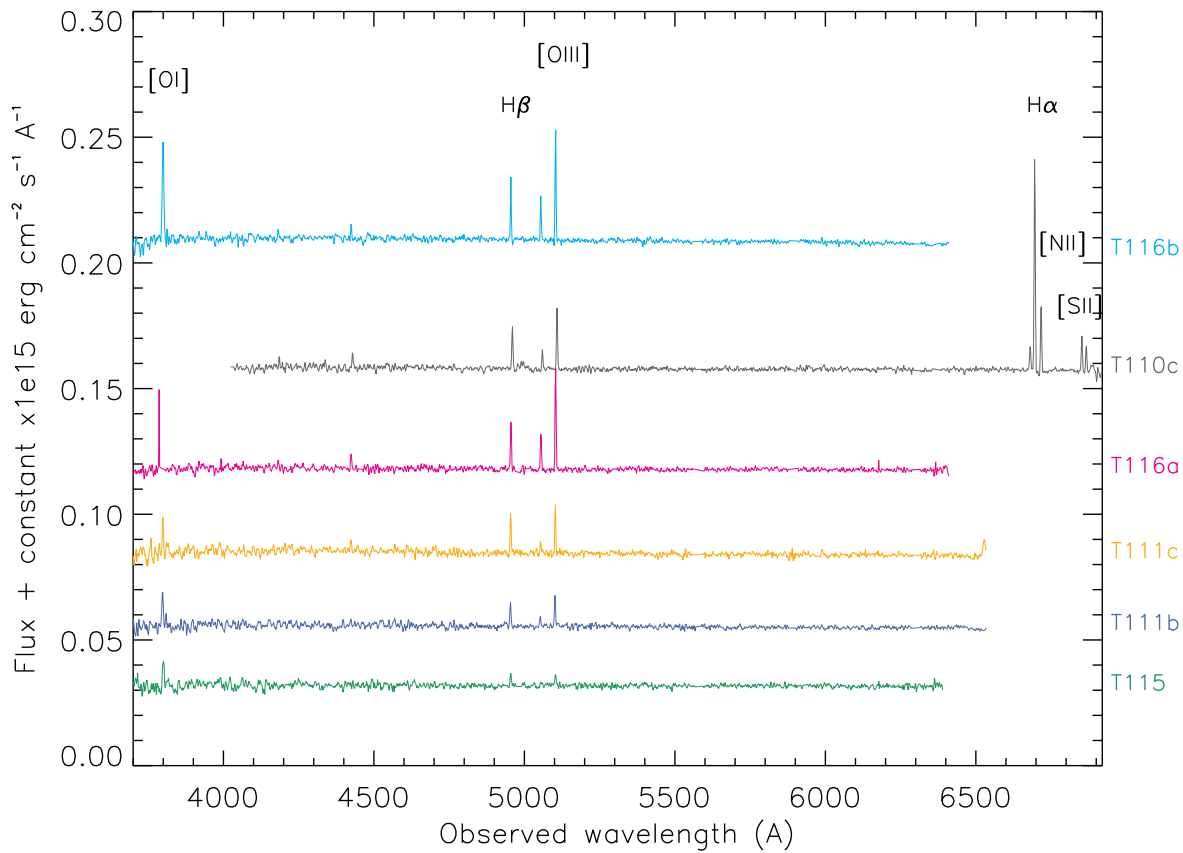}
 	\caption{Emission line spectra of seven star forming knots. We have ordered the spectra according to the strength of the H$\beta$ line, for illustration purposes. This is the lower tier of brightness, with the remaining emission line spectra presented in Figure~\ref {fig:spec3}. In all, no spectrum, of the ones where the H$\alpha$/N[II] region is covered, shows emission line ratios characteristic of shocks.  
			}\label{fig:spec2}

 \end{center} 
 \end{figure}

\begin{figure}
 \begin{center}
 	\epsscale{1.0}
 	\plotone{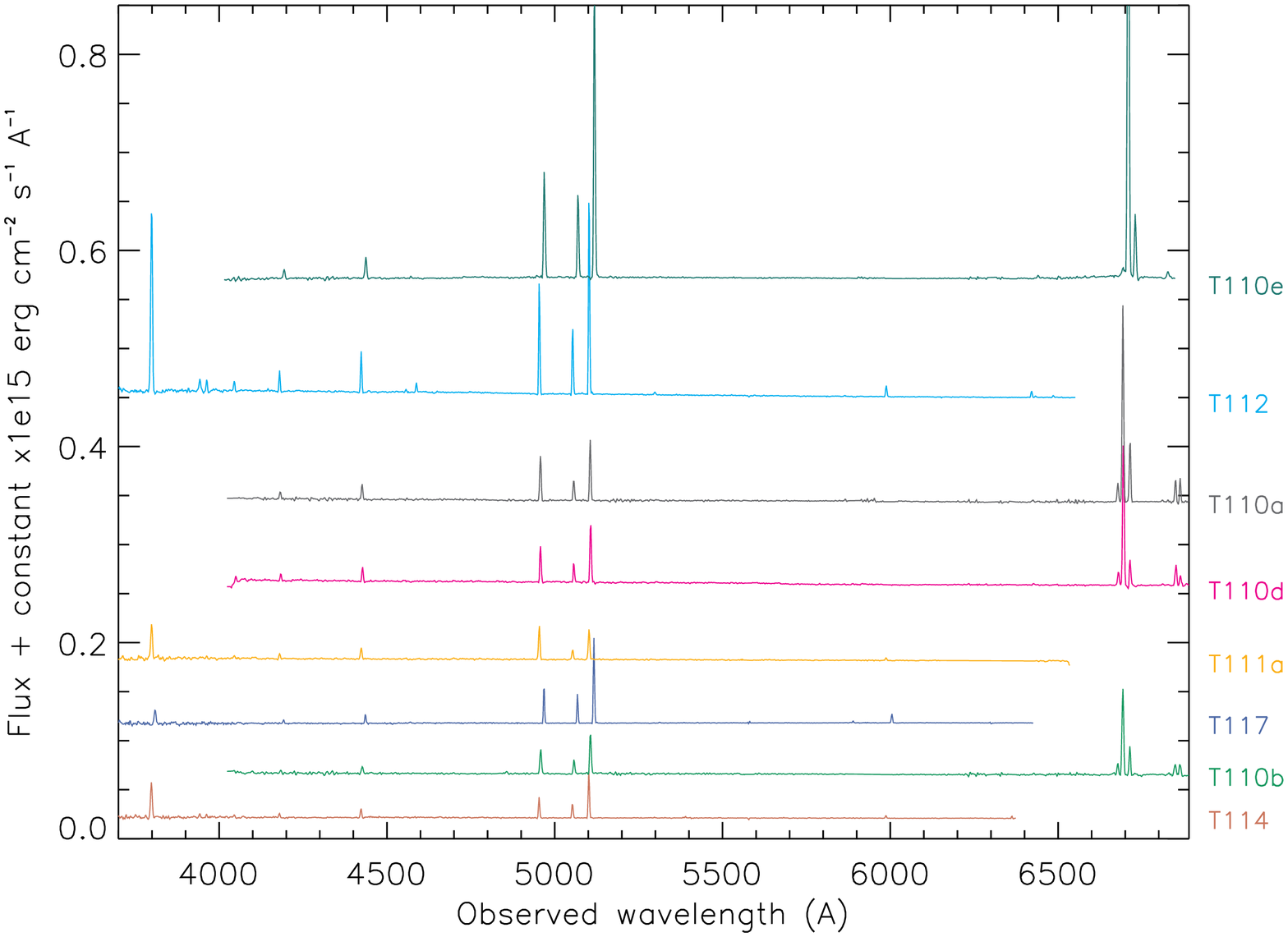}
 	\caption{Emission line spectra of a further seven star forming knots, continuing from Figure~\ref{fig:spec2}. All knots show H{\,\sc ii}~region-like emission line ratios. 
			}\label{fig:spec3}

 \end{center} 
 \end{figure}

{ The spectra are presented in the original binning and with no flux scaling in Figures~\ref{fig:spec1}~through~\ref{fig:spec3}. The strengths of the H$\beta$ and ~\Oiii~ as well as H$\alpha$, \Sii~ and \Nii~ where available, were measured and are presented in Table~\ref{table:specphot}.}

\subsection{The nature of emission-line objects}\label{sec:e-spec}
Given the location of some of the studied sources with respect to the strong X-ray component of SQ, it is crucial to understand the nature of our targets before proceeding with the analysis. The dust maps presented by Natale~et~al.~(2010) show a correlation between dust and X-ray emission in various parts of the Quintet, including the Northern Starburst Region (the SQ-A X-ray region), where one of our spectroscopic slits was placed (ID T110).  In addition, slits T119, T120, and T121 are projected on regions of X-ray emission. We find no striking evidence of shocks in any of the spectra. To confirm that, we employ three diagnostics to distinguish between H{\sc ii} regions and shock-ionized regions:  radial velocities, line broadening, and emission line ratios.

First, we make use of the radial velocities of the knots in question, as listed in Table~\ref{table:properties2}. The knots in slit T110 have velocities at the mid- and high-end of the range present in SQ ($\gtrsim6000~$\kms), therefore placing them potentially within or behind the shock region. Slits 119 through 121 register smaller velocities, which suggest they are in front of the shock along our line of sight, and therefore not surrounded by a shocked medium. Emission lines can be found in the sky dimension of our Gemini multi-extension spectra, and we use those to place the background emission in the complex velocity space of SQ. For example, in the subtracted sky of T121, we find a line consistent with \Oiiiu~ shifted to $\simeq5100~$\AA. This would place it at the same distance as the shocked medium, but not the knot targeted within the slit.  

We investigate further by measuring the width of the \Oiiiu~ line, which is the only line that is consistently covered in all spectra and one that could be broadened by shocks.  The resolution of our spectra varies between $3.5-4.4~$\AA, corresponding to velocities of $90-110~$\kms. Line broadening by shocks can only be inferred by values above the instrumental plus the associated error. Comparing the width of the \Oiiiu~ line to the instrumental width, as measured from a sky line, reveals no significant broadening in any of the spectra. More specifically, the quadratic difference (an adequate measure of broadening) of the FWHM of \Oiiiu ~minus that of a sky-line, 

$[(\textup{FWHM}_{\textup{\scriptsize\Oiiiu}})^2 - 
	(\textup{FWHM}_{\textup{\scriptsize sky}})^2]^{-2}$
	
typically registers at $\simeq65\pm32$ km\,s$^{-1}$ (sample median and standard deviation), which is not large enough to claim bona-fide broadening. 

\begin{figure}
 \begin{center}
 	\epsscale{1.0}
 	\plotone{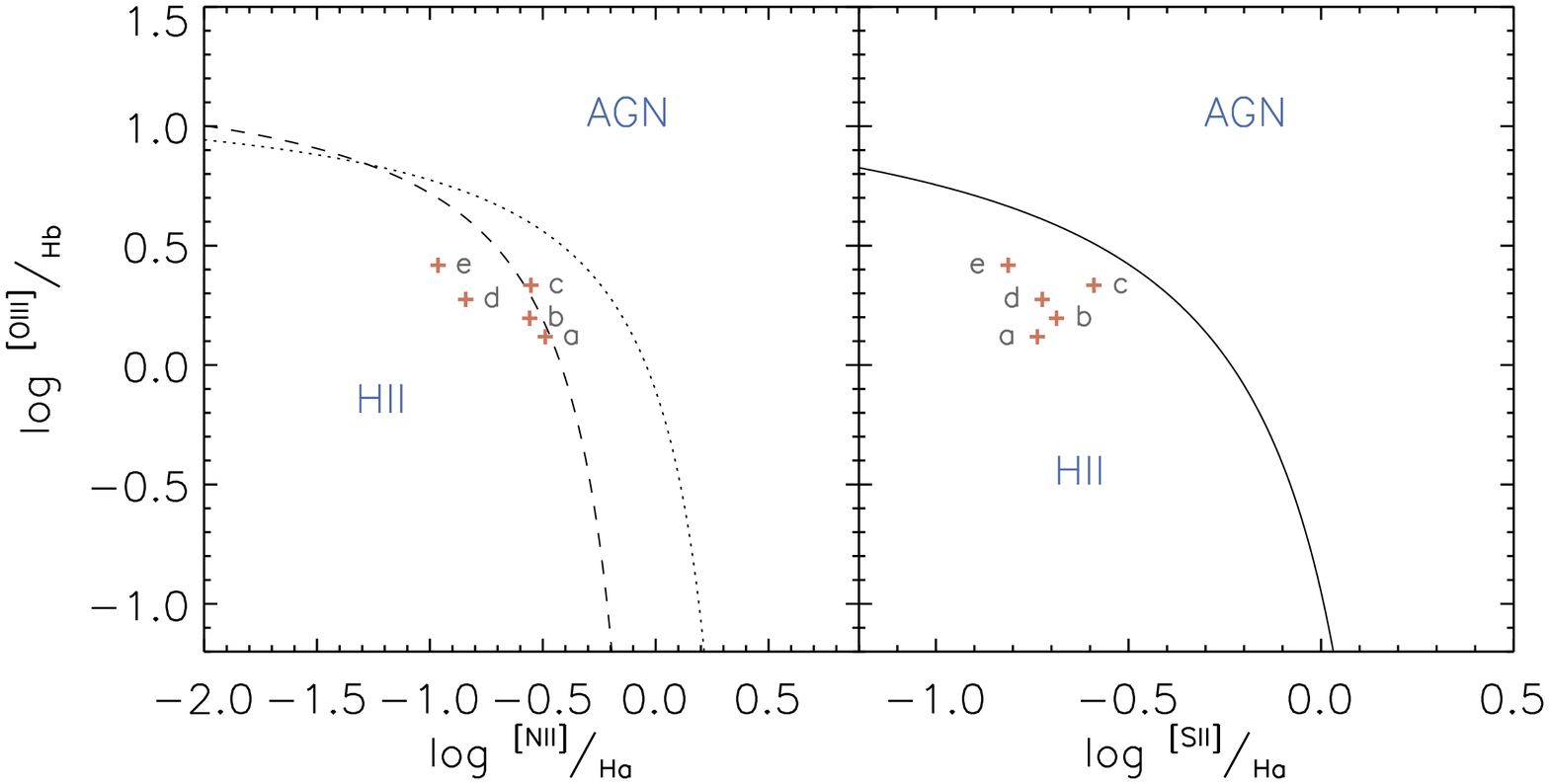}
 	\caption{ `BPT' diagram with the \hii/AGN diagnostic line of 
				Kaufmann~et~al.~(2003). We plot the following ratios of 
				\Oiii/H$\beta$ versus \Nii/H$\alpha$ and \Oiii/H$\beta$ 
				versus \Sii/H$\alpha$  available in slit T110 that contains
				five spectra of H$\alpha$-bright knots in the Northern 
				Starburst Region. Lettering refers to the specific ID of each 
				knot within slit T110. The knots all register below the AGN line, therefore
				indicating the absence of shocks. While the radial velocities
				of these knots place them within or behind the shocked medium 
				detected in X-rays (see section \ref{sec:e-spec}), the placement of sky-subtraction
				apertures locally within each slit keeps the spectra free of 
				contamination by the surrounding emission. 
			}\label{fig:bpt}

 \end{center} 
 \end{figure}

We then measure the different ratios of \Oiii/H$\beta$, \Nii/H$\alpha$ and \Sii/H$\alpha$ available only in slit T110. This metric is nearly independent of extinction, as it employs ratios of neighboring lines. All ratios qualify the knots as \hii~regions, as diagnosed by comparing to the `BPT' diagram of Kewley~et~al.~(2006; based on \citealt{bpt}).  We make use of the Kaufmann et al. (2003) diagnostic of line ratios that delimits the parameter space representative of \hii~ regions, as shown in Figure~\ref{fig:bpt}. None of the knots lie on the AGN side of the Kaufmann~et~al. line. In conclusion,  we have no evidence of shocked gas in our spectra. 

Finally, comparing the ``background" spectrum to the ``on-target" spectrum for each source shows that the diffuse component is much fainter than the target source.  Hence, any diffuse component (from shocks or nearby star-formation) does not significantly contribute to the observed spectrum.  We conclude that the presented spectroscopy is not affected by any surrounding ionized material.
 
\subsection{HST photometry}
\label{sec:photometry}

{\em HST}/WFC3  F438W, F606W, F814W, and F657N images of the SQ were taken from Early Release Science archive.  The images came fully reduced and drizzled by the standard automatic pipeline.    See F11 for more information regarding the images. Photometry was performed on the images with apertures the same size as the slit width in order to directly compare the spectroscopic and photometric results.  The aperture for photometry was 10~pixels (165~pc) in radius and the inner/outer background radii were 11~pixels (181~pc) and 13~pixels (214~pc) respectively. The numbers in parentheses are the projected physical sizes assuming a distance to the system of 87.1~Mpc (adopting $H_{0} = 70$~km\,s$^{-1}\,$Mpc$^{-1}$).   

\begin{figure*}
 \begin{center}
   \epsscale{0.9}
\includegraphics[width=3.4cm]{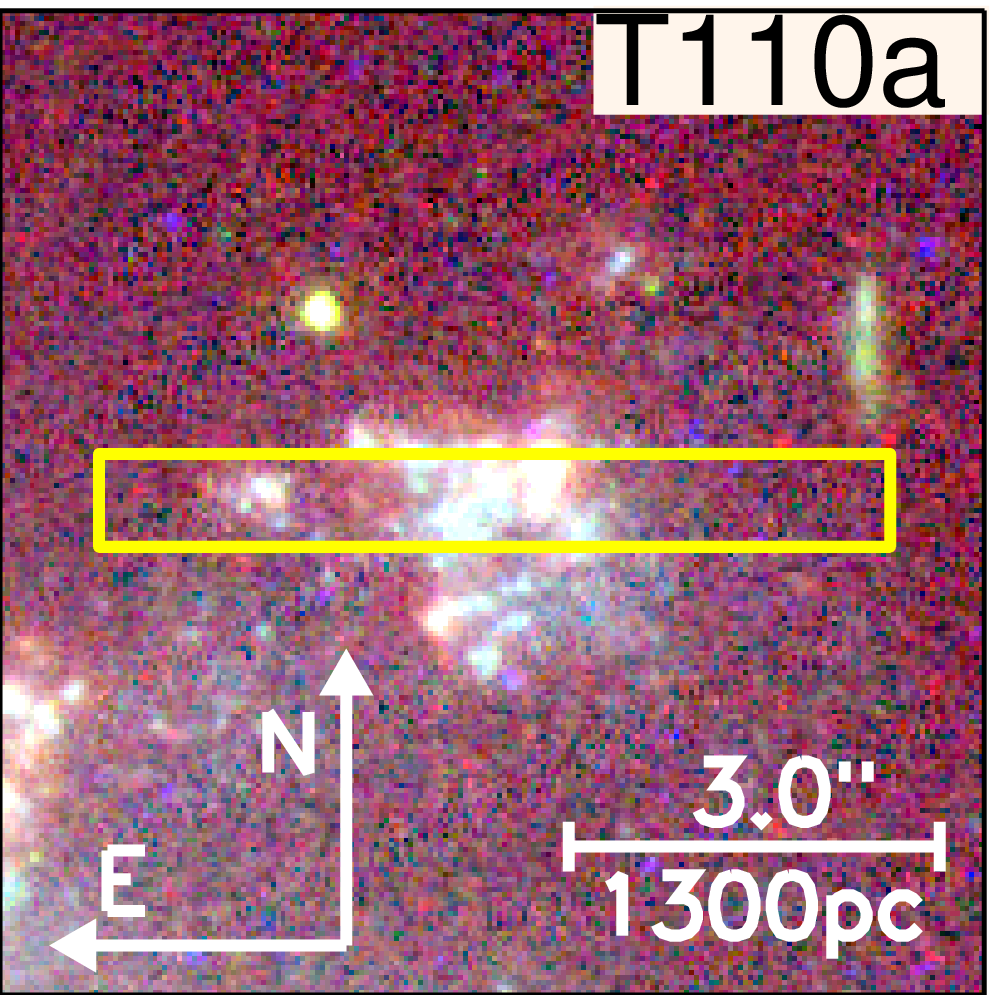}
\includegraphics[width=3.4cm]{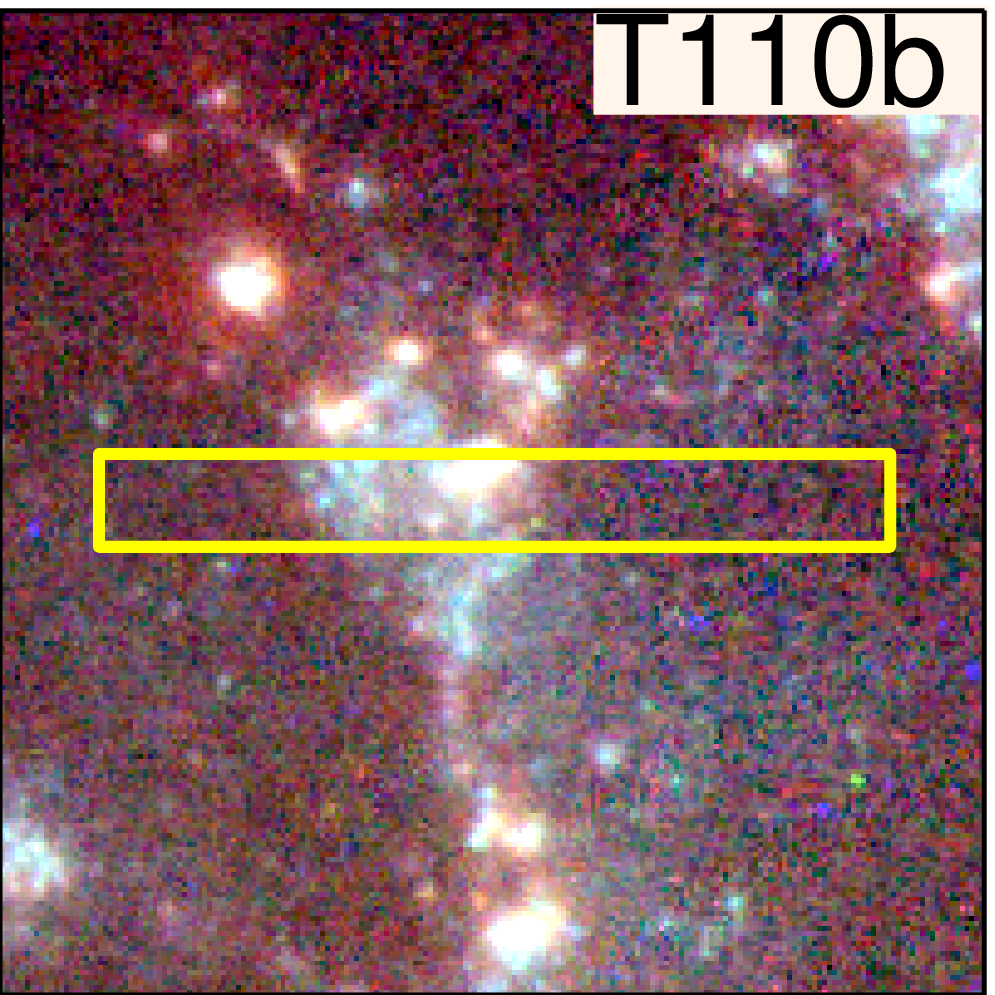}
\includegraphics[width=3.4cm]{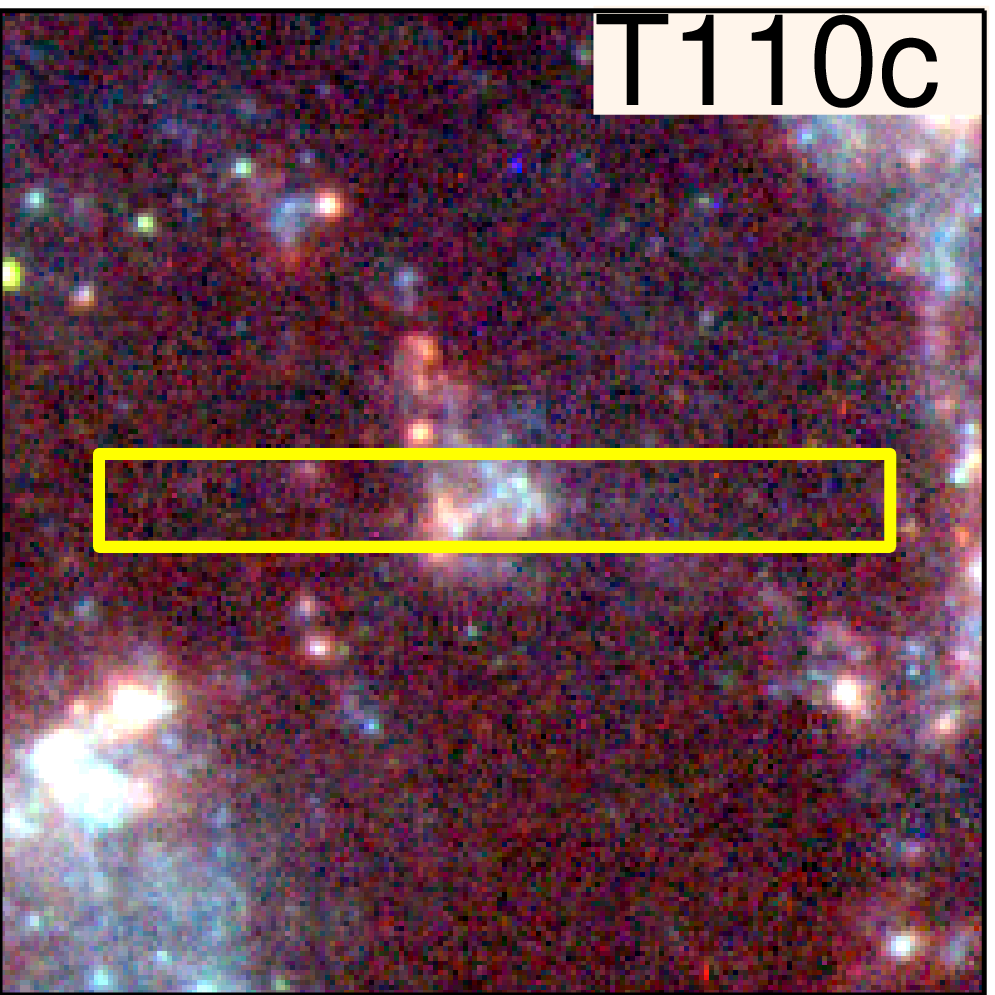}
\includegraphics[width=3.4cm]{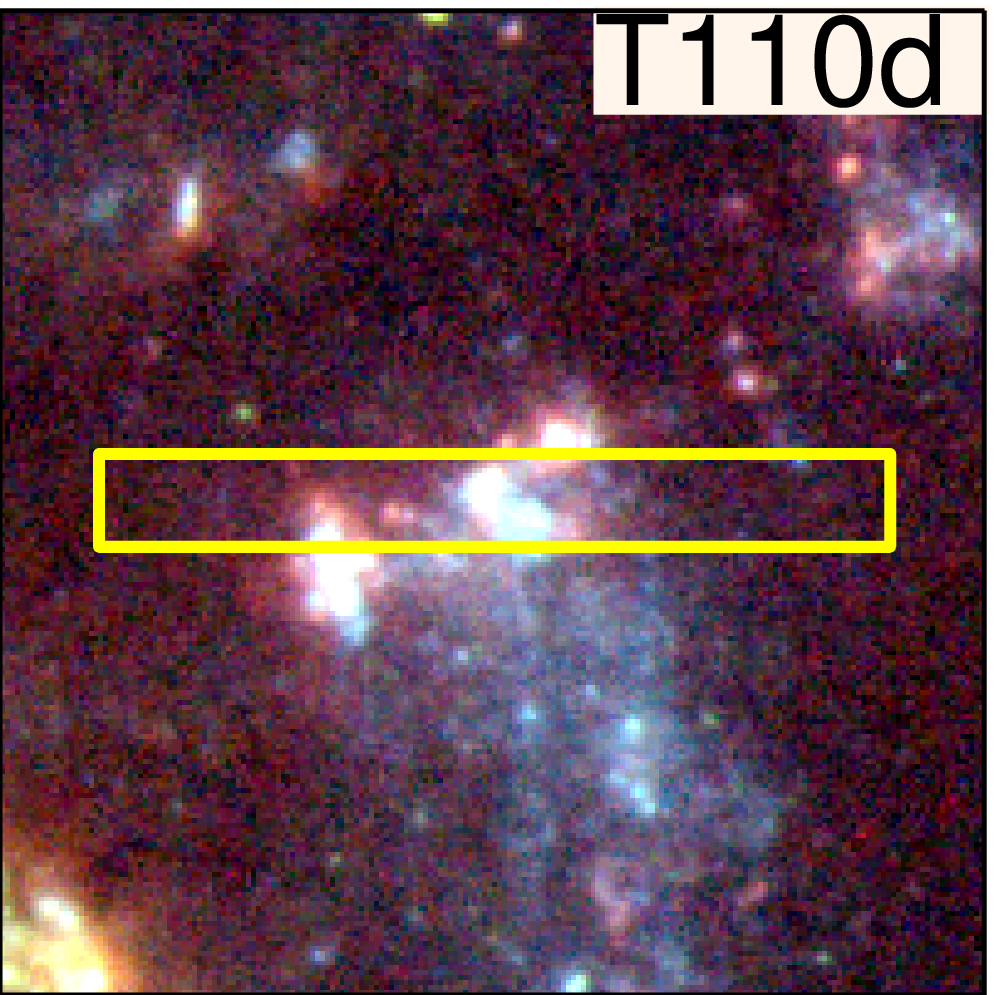}\\
\includegraphics[width=3.4cm]{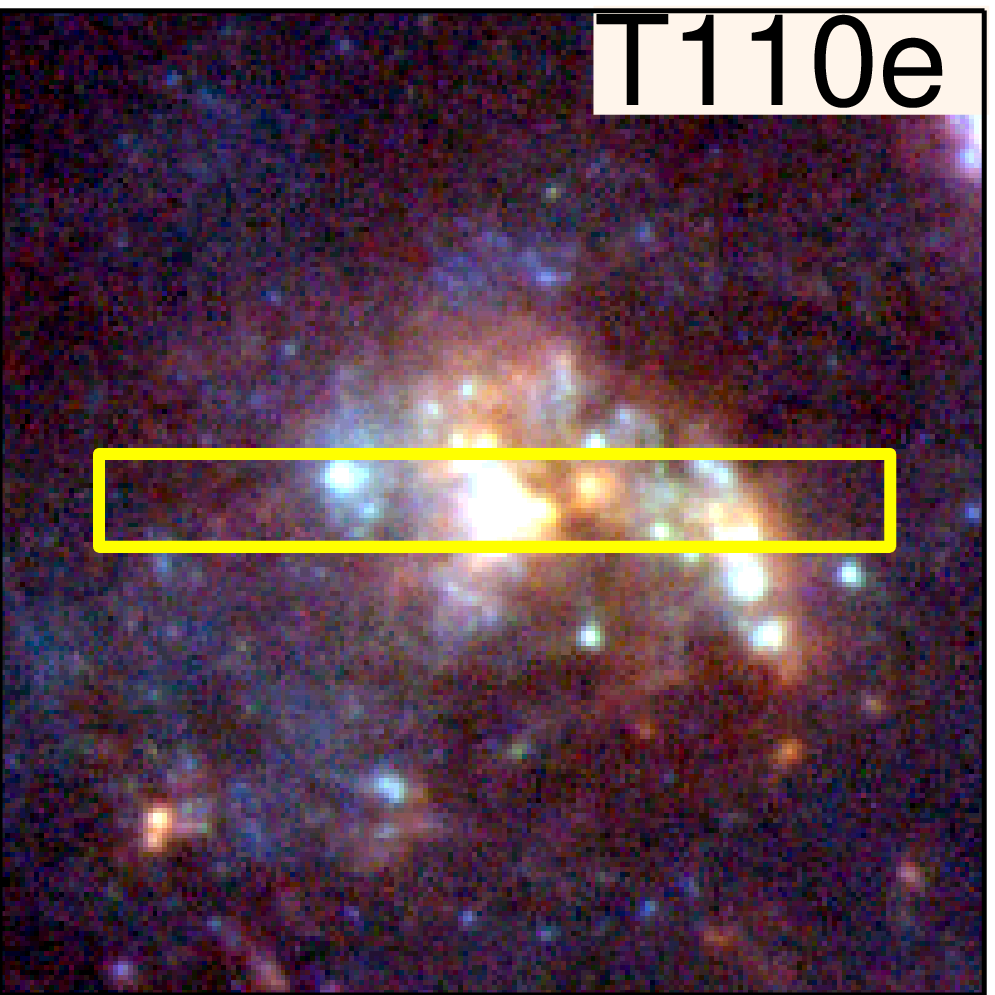}
\includegraphics[width=3.4cm]{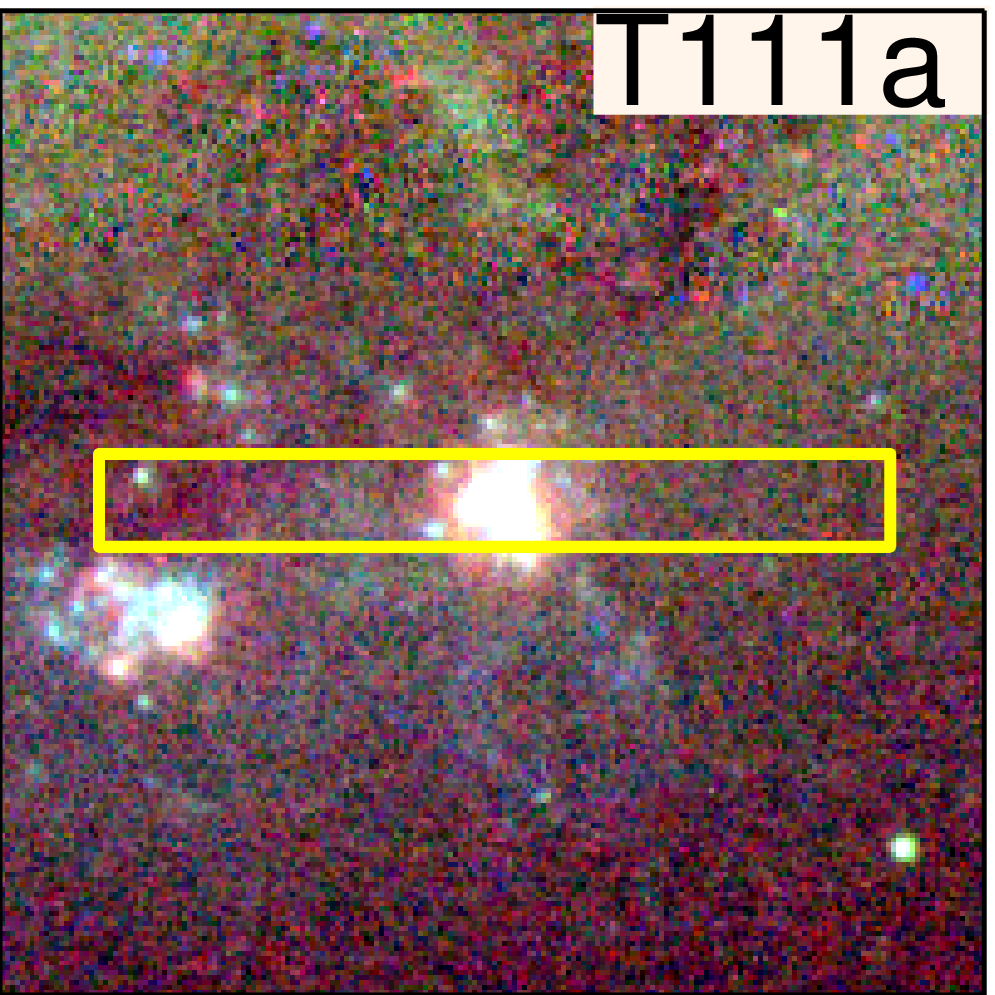}
\includegraphics[width=3.4cm]{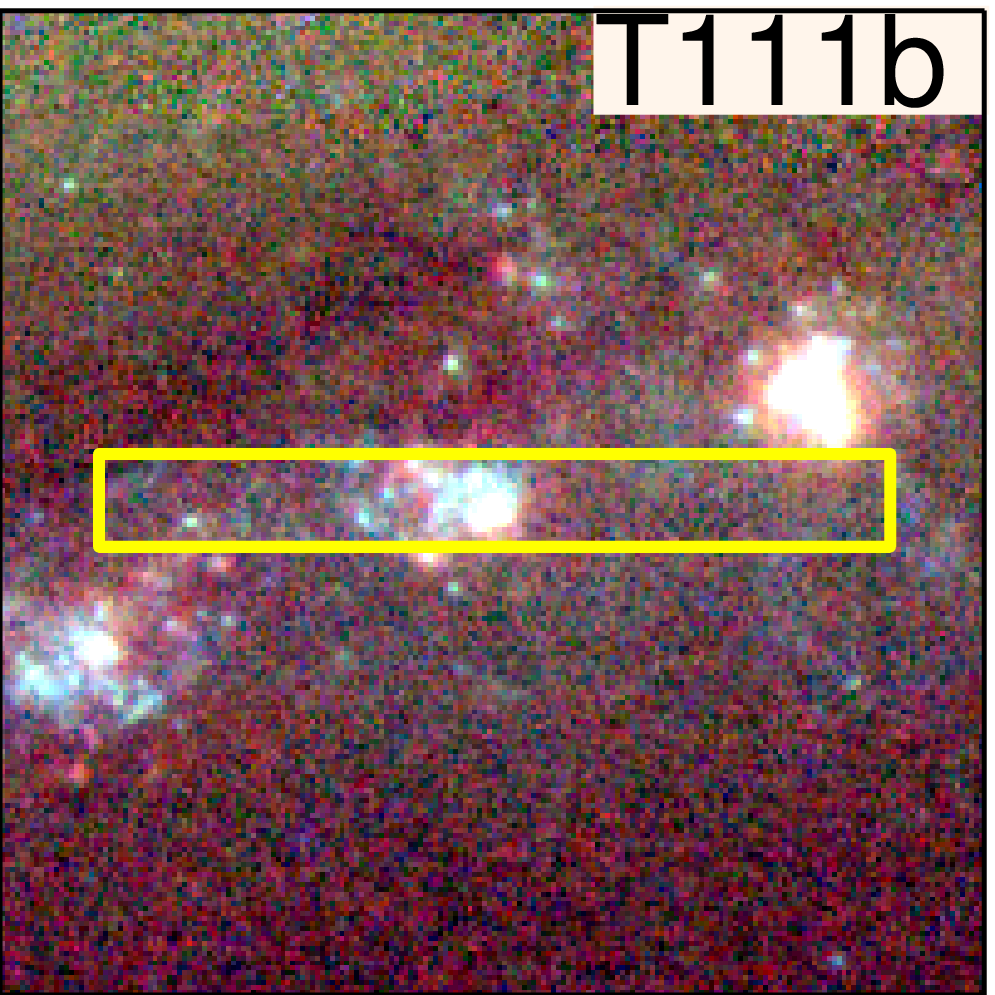}\\
\includegraphics[width=3.4cm]{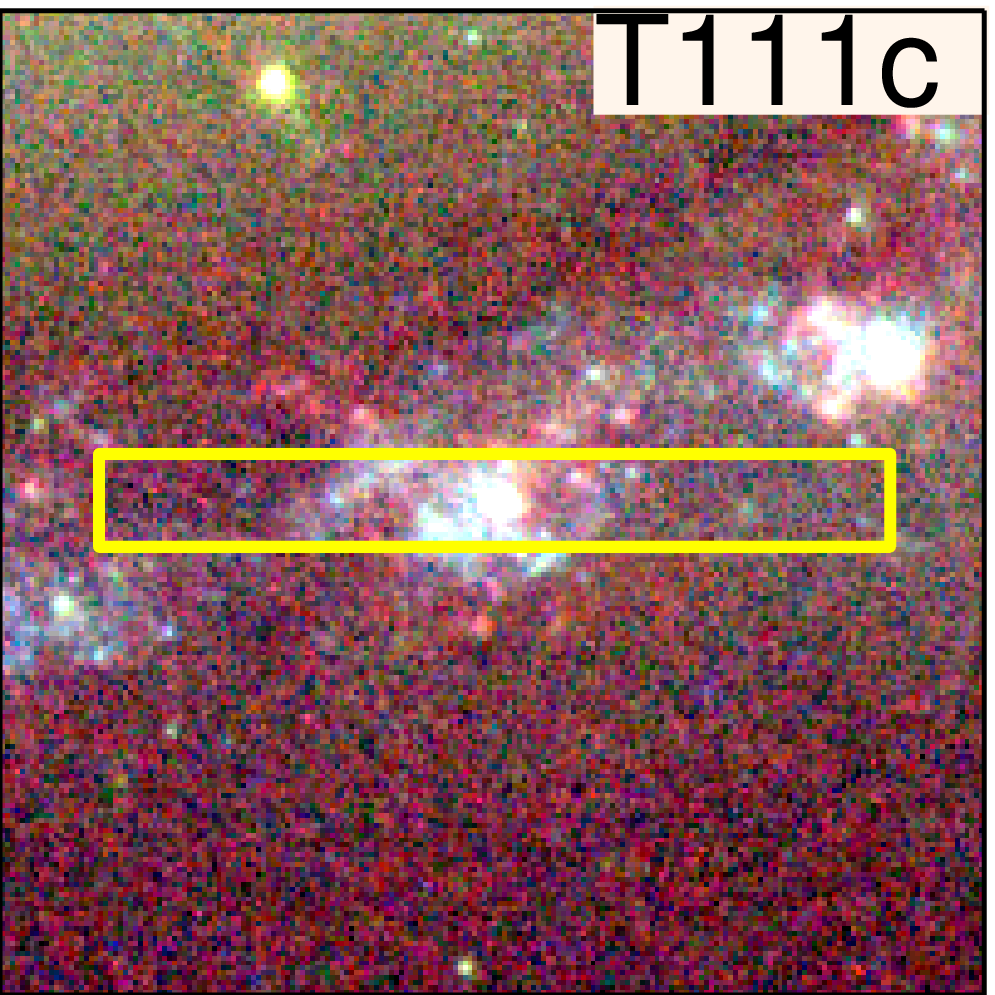}
\includegraphics[width=3.4cm]{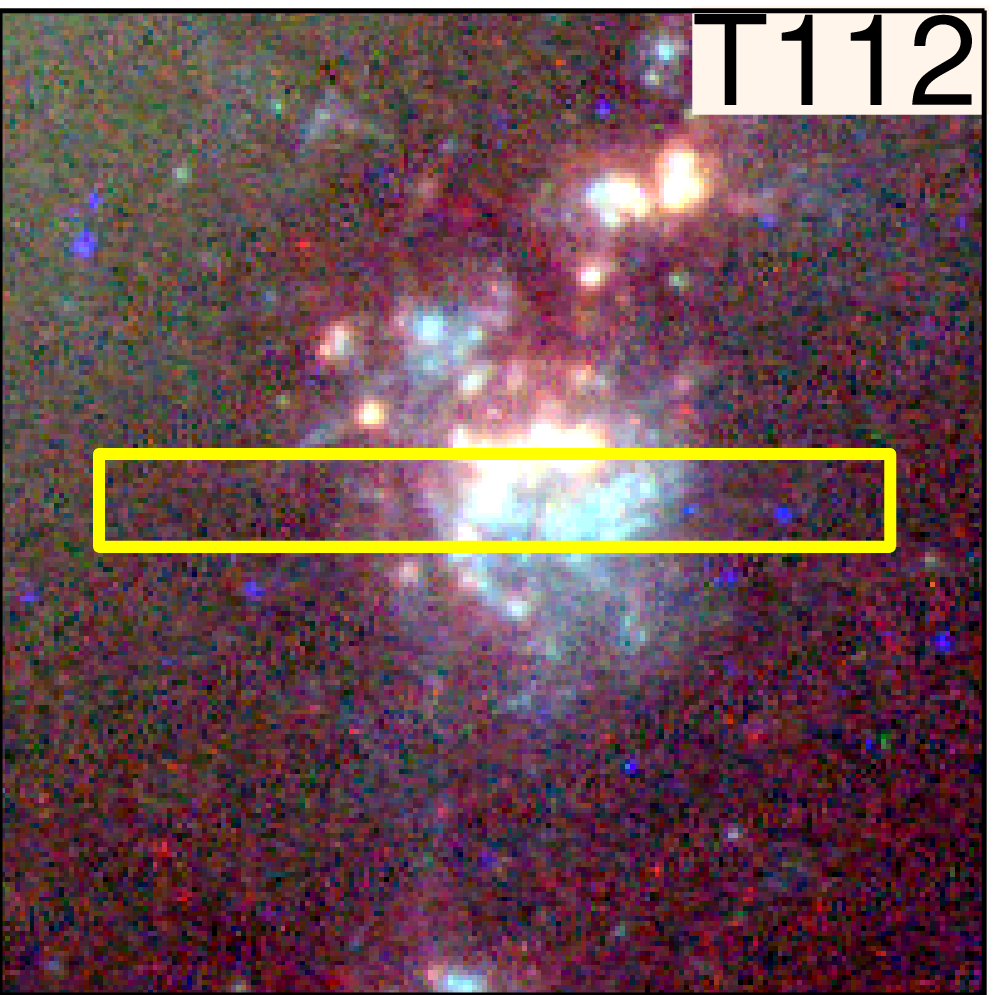}
\includegraphics[width=3.4cm]{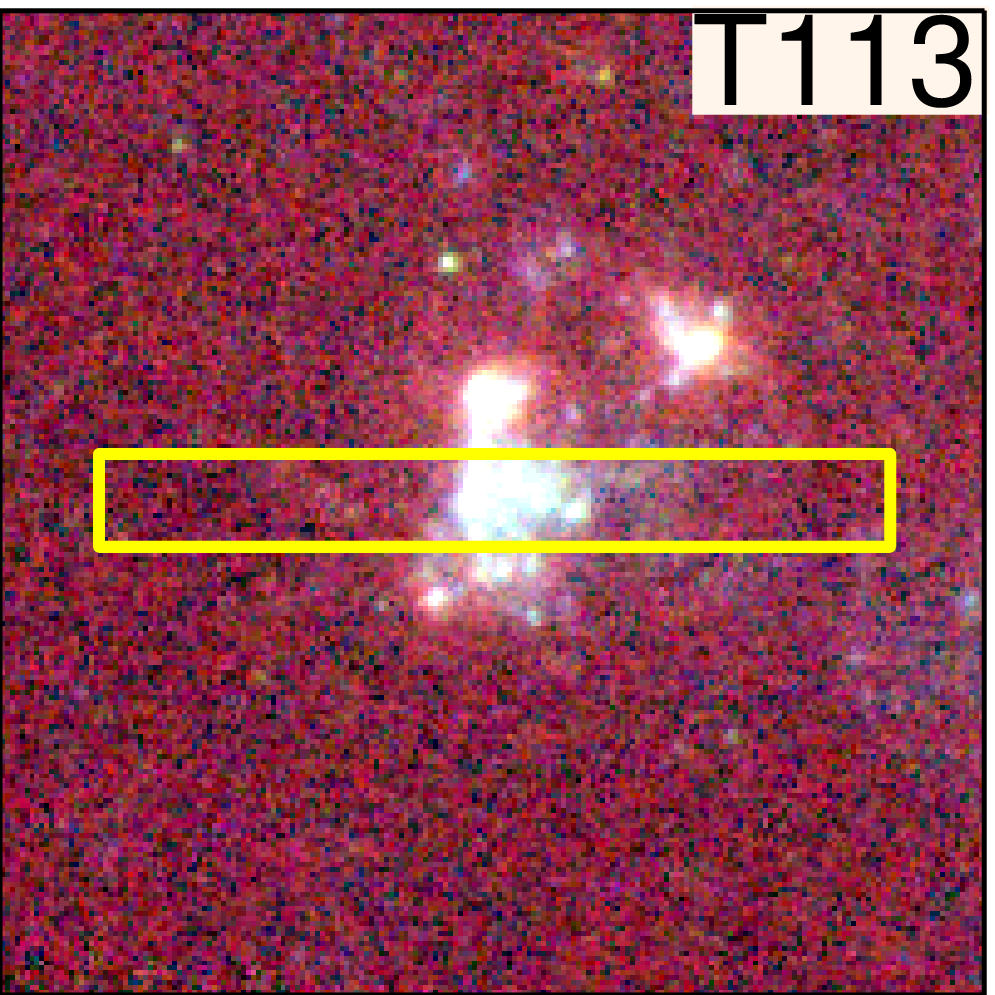}
\includegraphics[width=3.4cm]{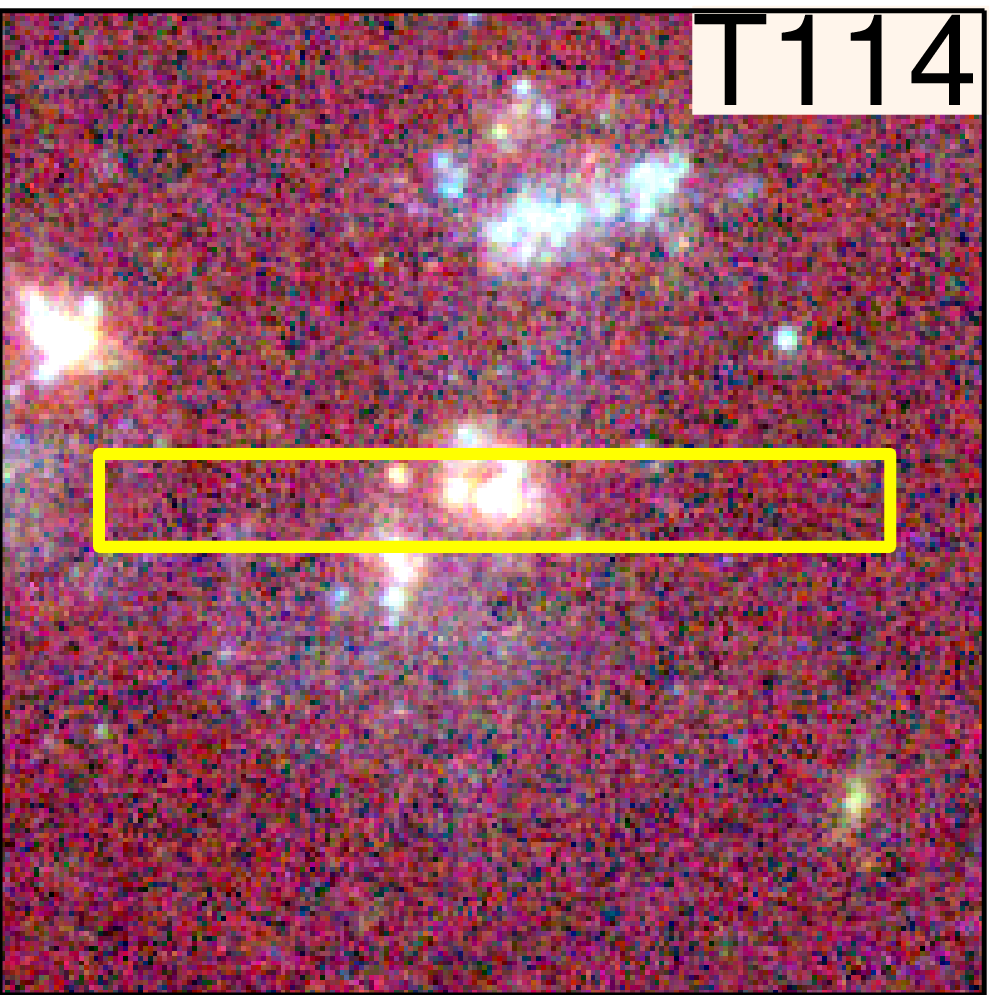}\\
\includegraphics[width=3.4cm]{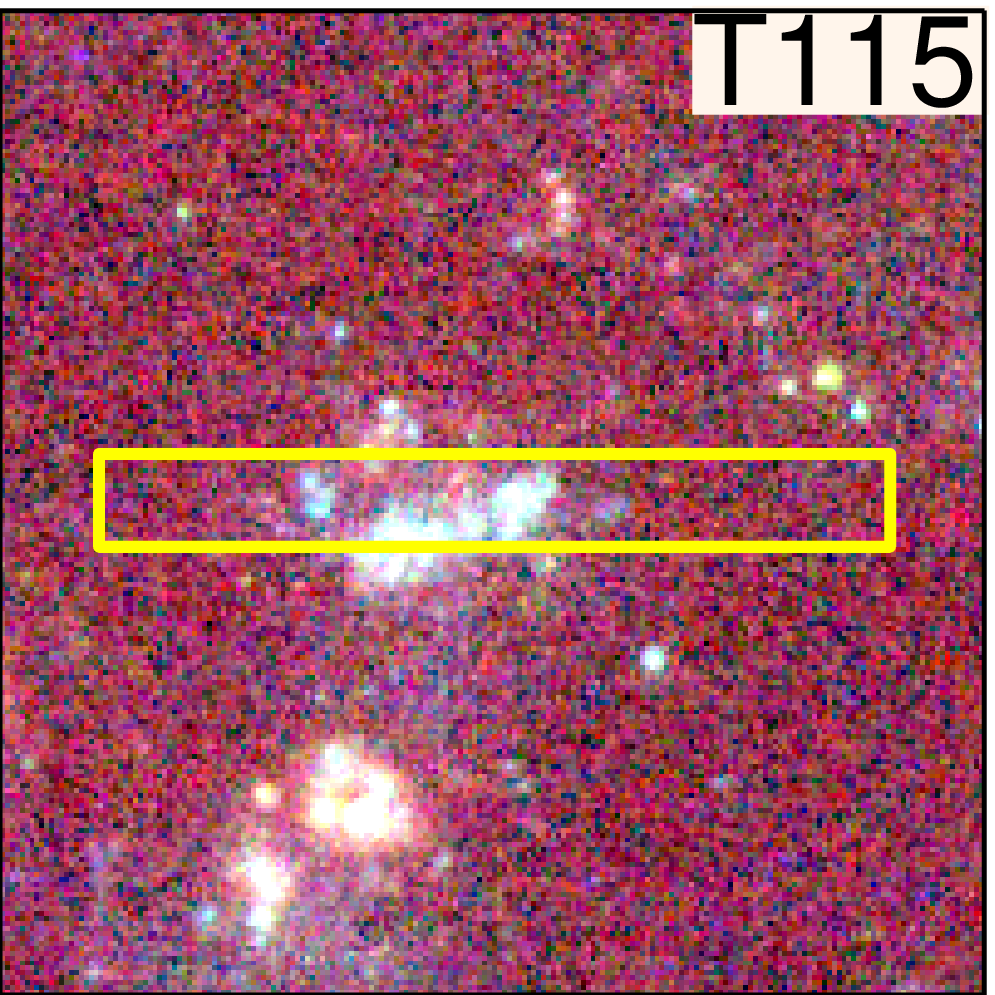}
\includegraphics[width=3.4cm]{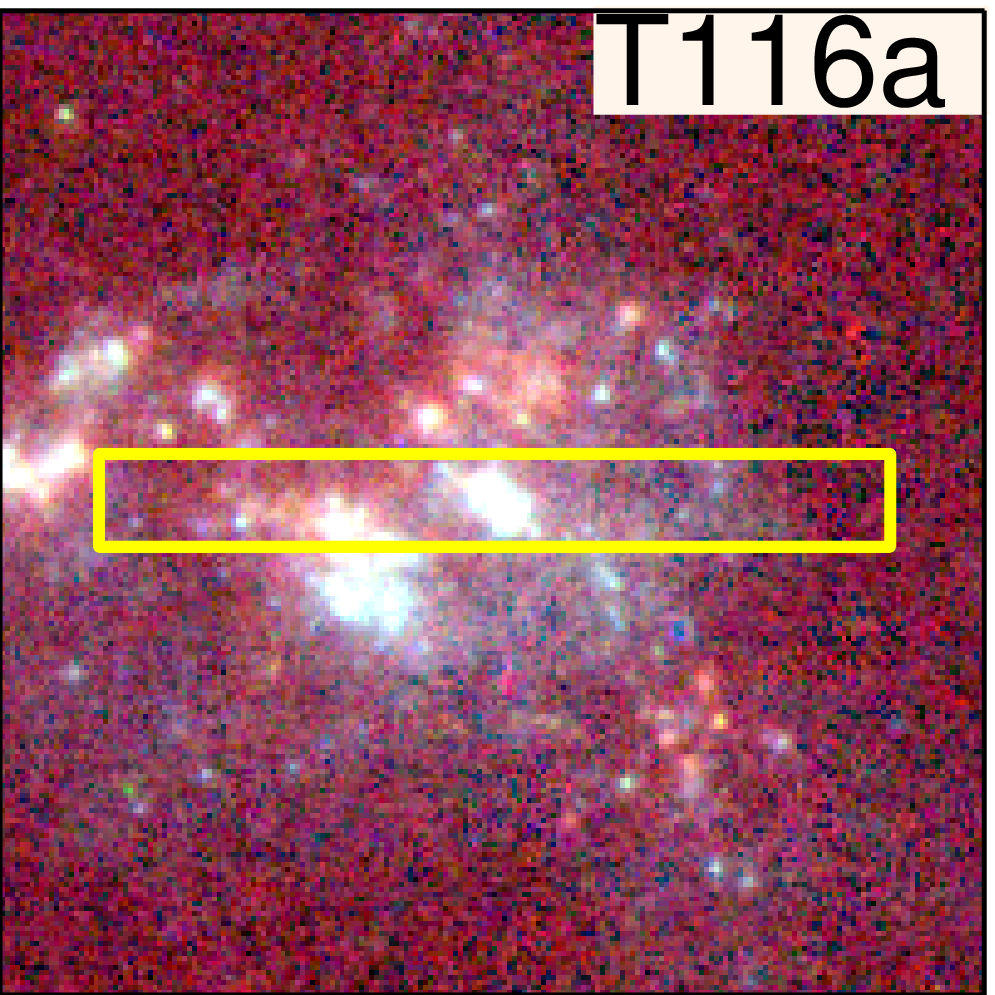}
\includegraphics[width=3.4cm]{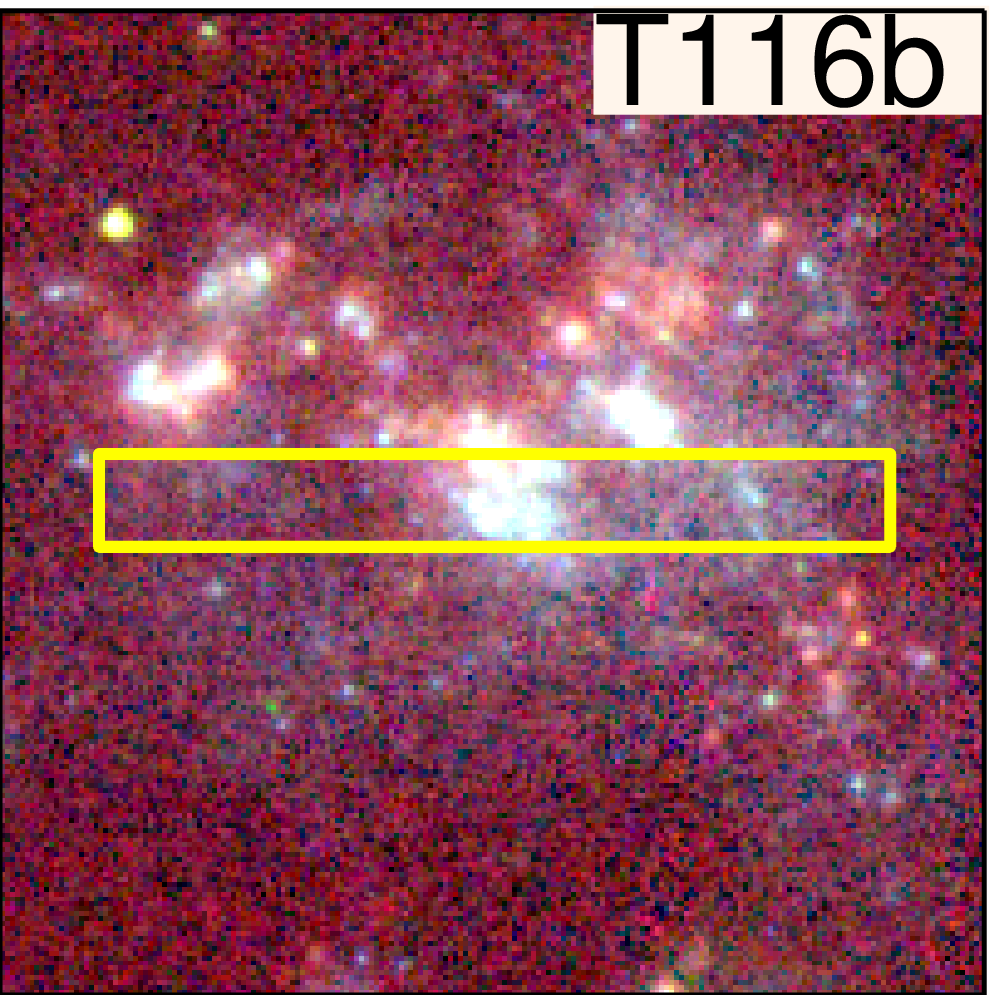}\\
\includegraphics[width=3.4cm]{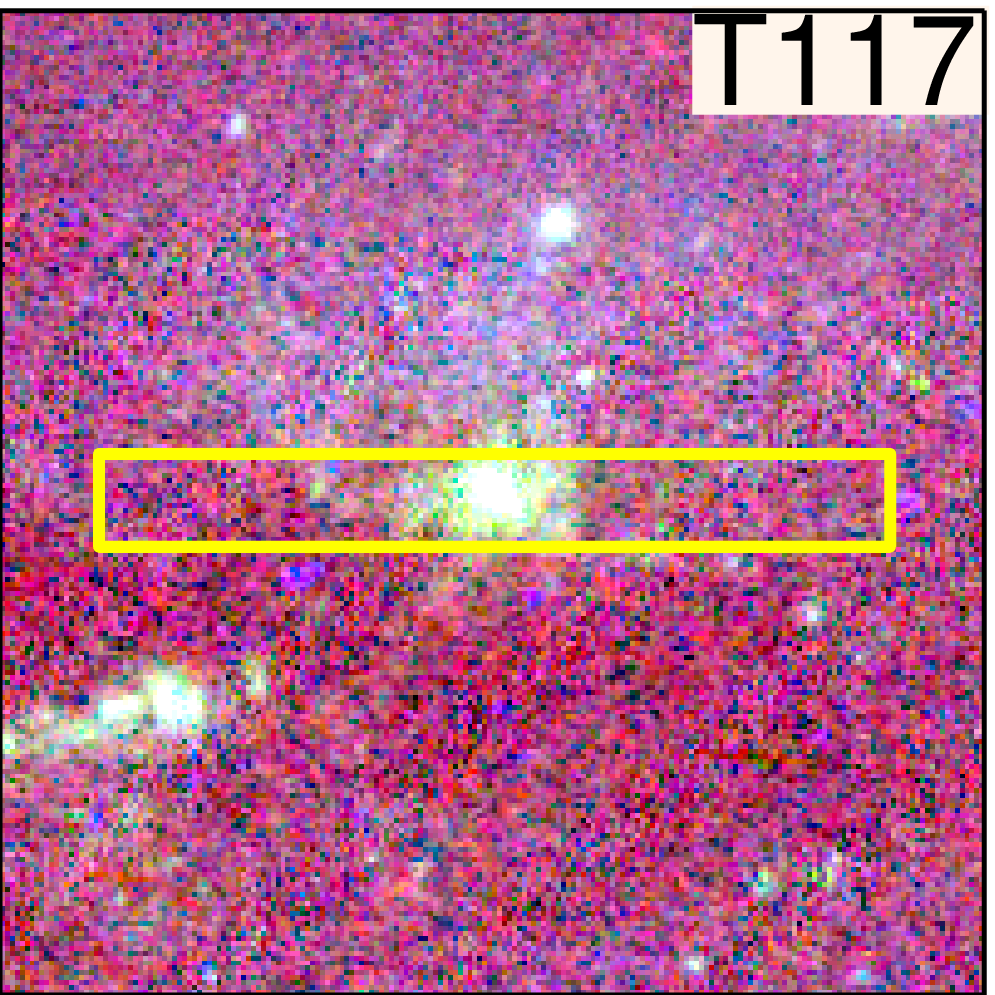}
\includegraphics[width=3.4cm]{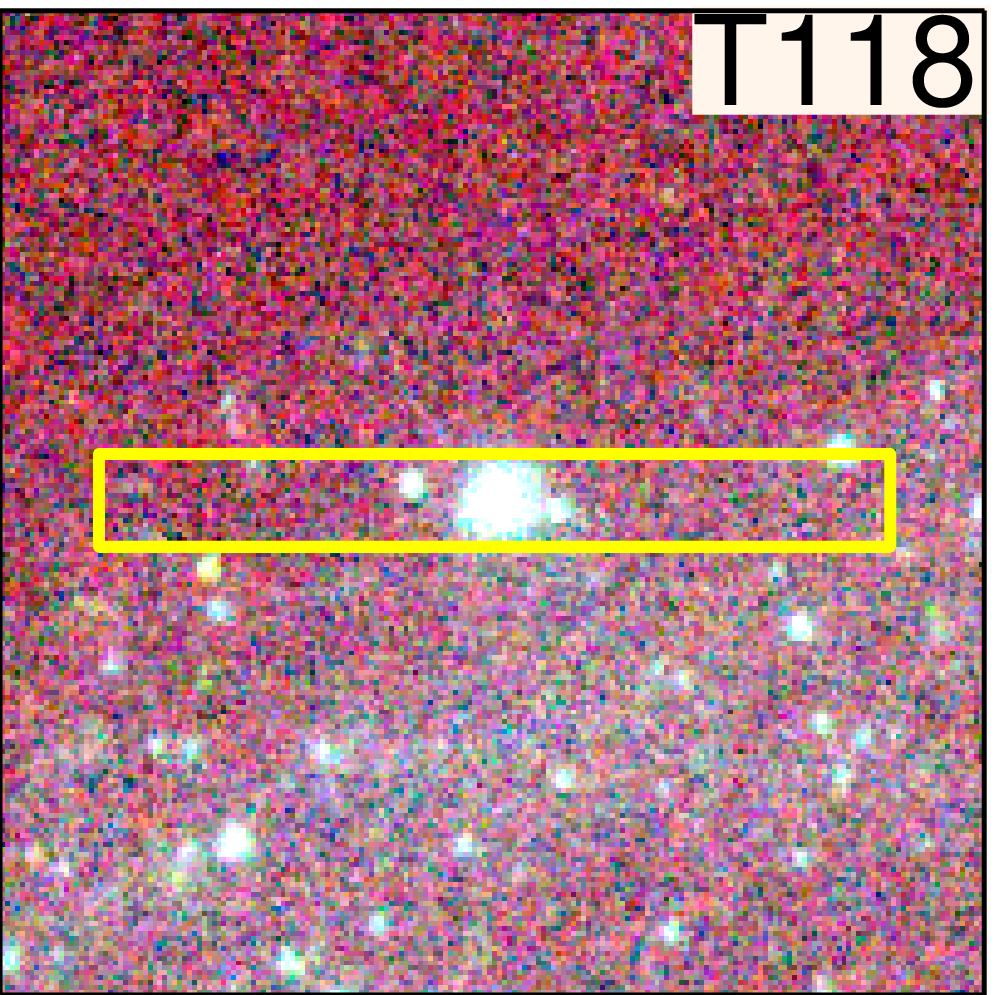}
\includegraphics[width=3.4cm]{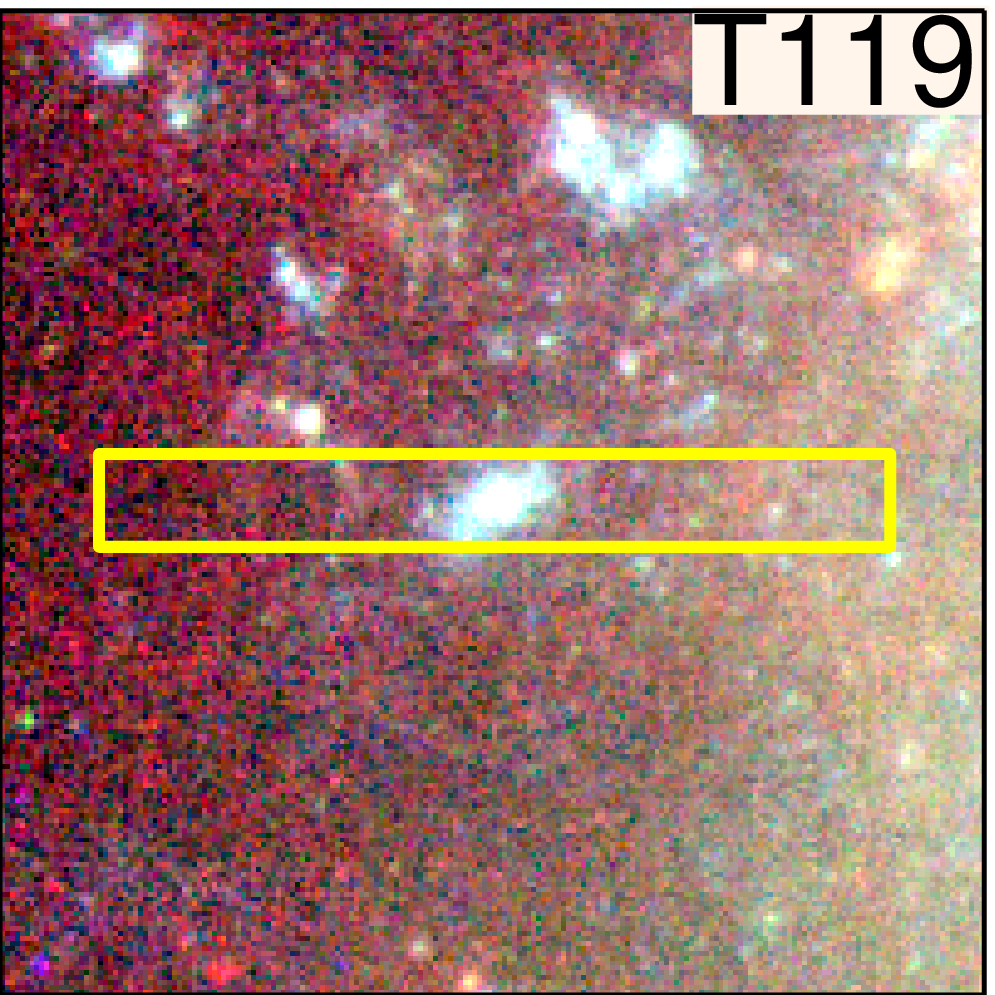}
\includegraphics[width=3.4cm]{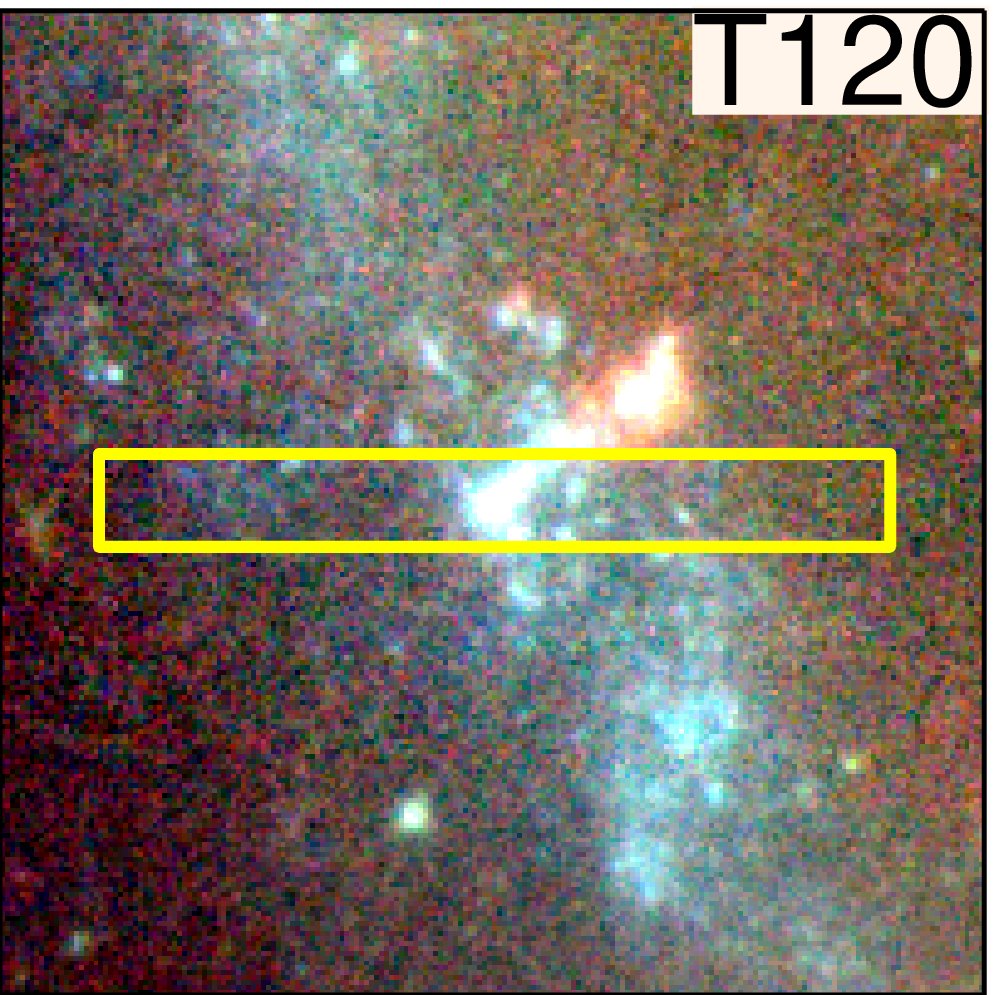}\\
\includegraphics[width=3.4cm]{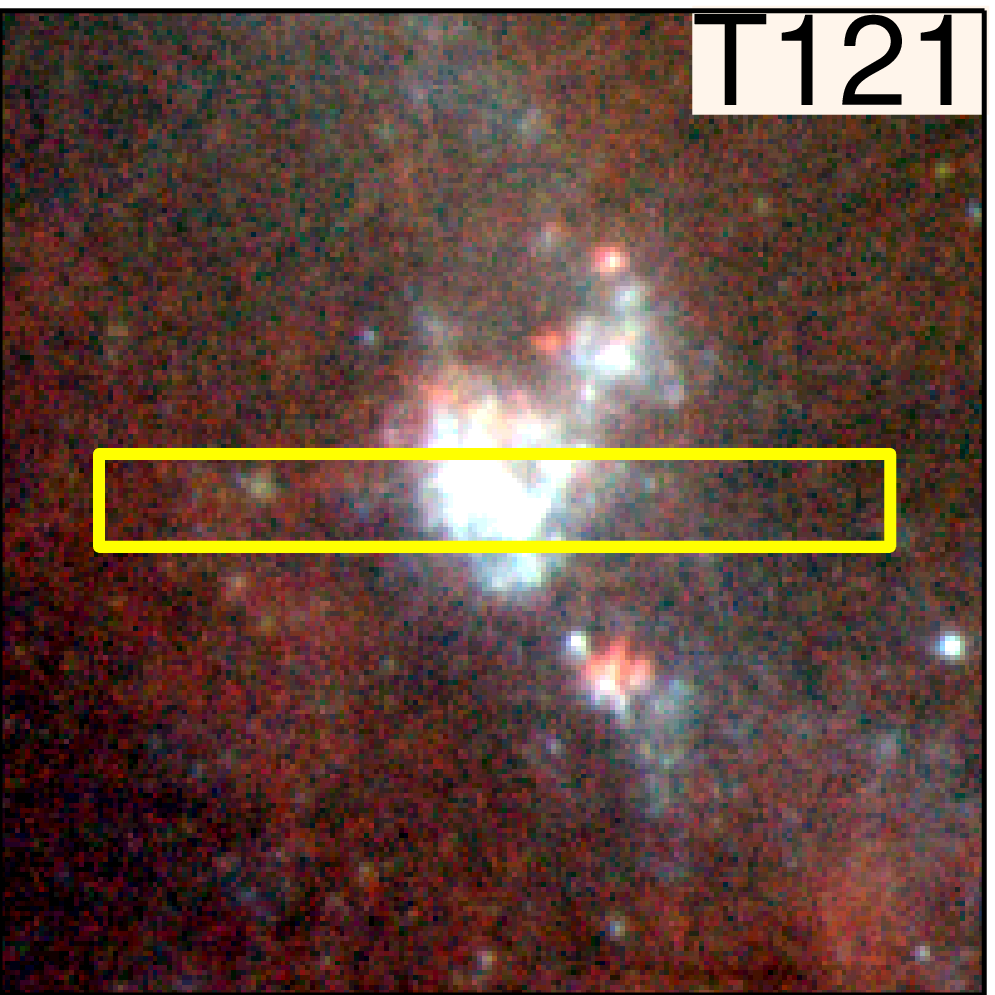}
\includegraphics[width=3.4cm]{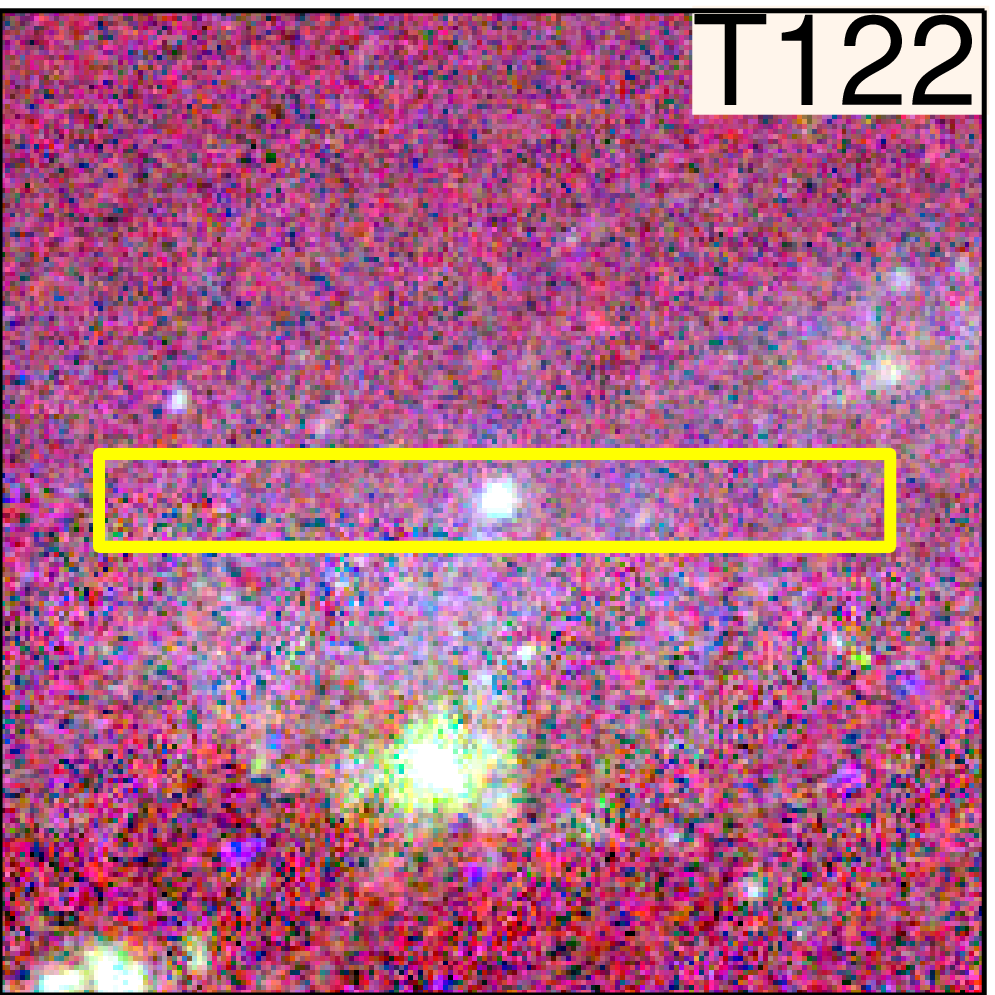}
\includegraphics[width=3.4cm]{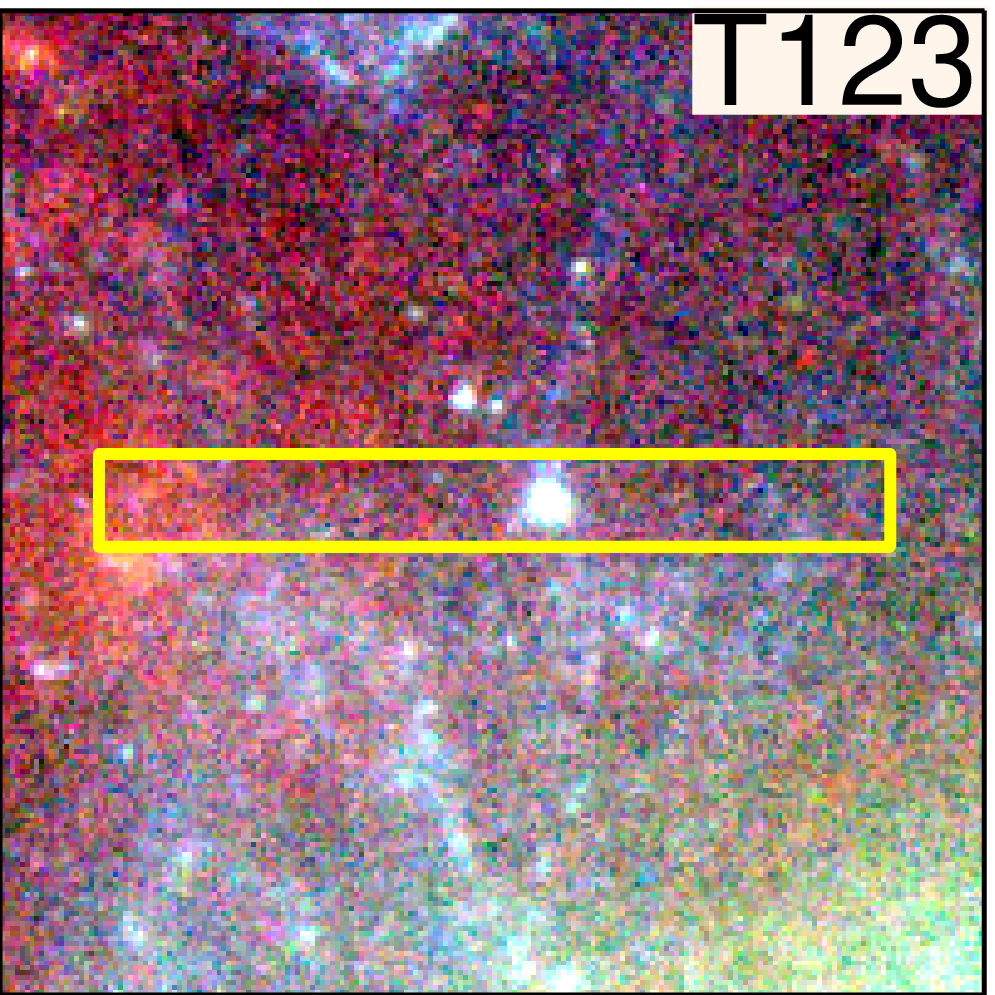}

	\caption{HST WFC3 B, V, and H$\alpha$ color composite images of the clusters/complexes presented in this study.  Each image is 200~pixels on a side, corresponding to $\sim3.3$~kpc at the adopted distance of 87.1~Mpc.} 
		\label{fig:images}
	\end{center} 
\end{figure*}

In Fig.~\ref{fig:images} we show color images centered on each of the observed clusters. The postage stamps measure 200~pixels, or$\sim3.3$~kpc, on a side.  The GMOS slit is shown in each image.  In Table~\ref{table:properties} we give the measured magnitudes for each cluster, where we have only corrected for Galactic foreground extinction (Schlegel et al. 1998).  Confirmed members and non-members are listed in Tables~\ref{table:properties2} \& \ref{table:comparison1}, respectively.

\begin{figure}
	\plotone{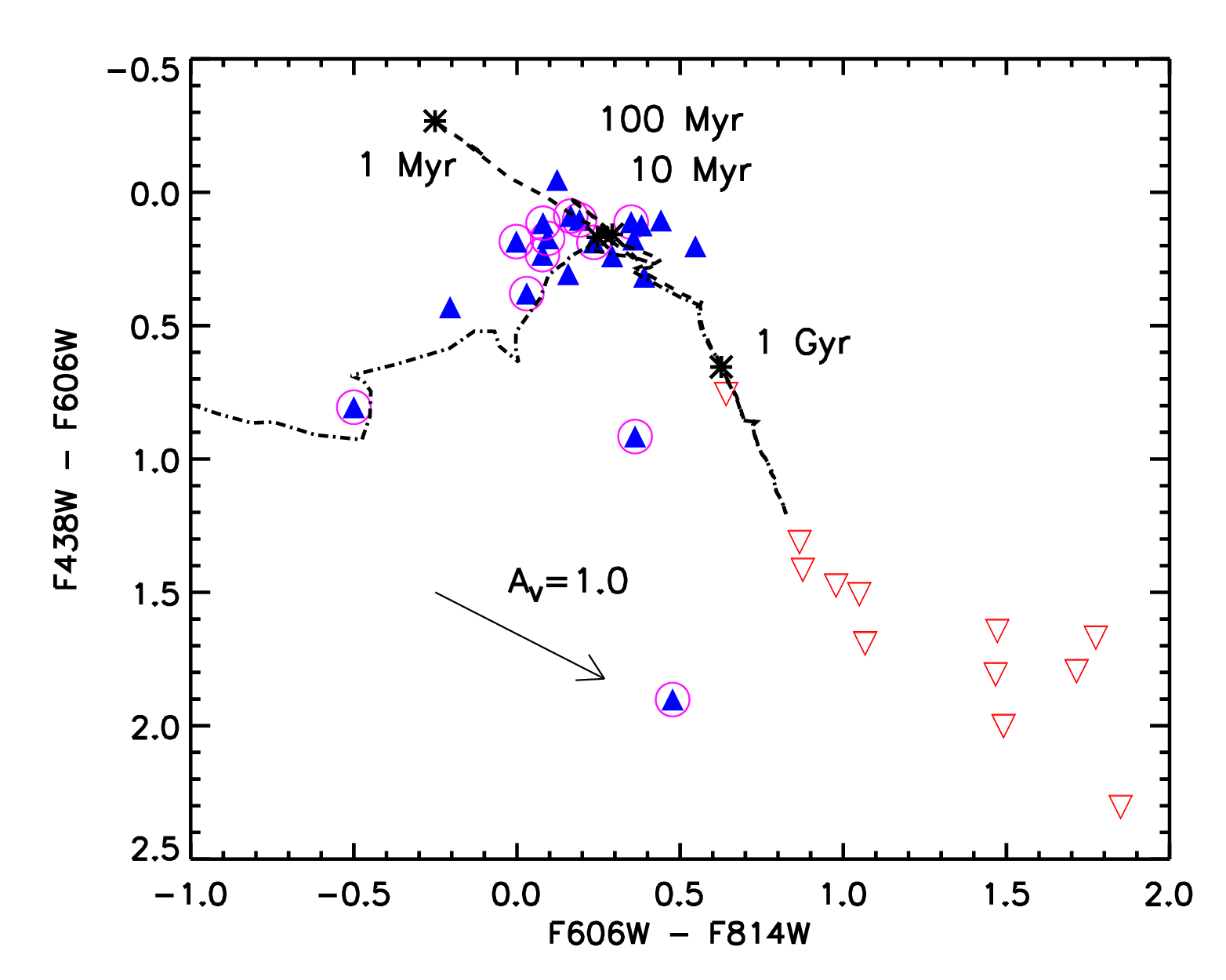}
	\caption{Color-color diagram of the targets presented in this work.  Filled (blue triangles) points represent clusters/complexes confirmed to be part of Stephan's Quintet.  Open (red upside down triangles) are targets that we confirmed not to be part of the system (either foreground stars or in one case, a background quasar).  Additionally, we show the Marigo et al.~(2008) simple stellar population (SSP) models (solar metallicity, IMF after Kroupa~1998) as a dashed line (stellar continuum only) and the asterisks mark  log age/yr = 6.6, 7.0, 8.0 and 9.0.  The dash-dotted line shows the Starburst99 SSP model tracks with nebular emission (Leitherer et al.~1999), { as in F11. The track spans the age range between 0.1 and 10~Myr, at which point it joins the regular track}.  Circled points denote sources with significant nebular emission associated with them as seen from the spectra. { Most of the spectroscopically confirmed clusters/complexes have locations in color space consistent with the model tracks. There are two exceptions, most notably T117 at (1.9, 0.5 coordinates), apart from which the confirmed clusters do not show evidence for significant reddening}. }
	\label{fig:cc}
	\end{figure}

Each of the targeted sources is shown in color-color space in Fig.~\ref{fig:cc}.  Filled (blue) triangles show sources confirmed to be part of the SQ system, while open upside-down (red) triangles denote sources that were found to lie at discordant redshifts, being either foreground stars or background quasars.  In addition, we show the Marigo et al.~(2008) simple stellar population models for solar metallicity as a dashed line and mark ages of 1, 10, 100, and 1000~Myr.  Some of the confirmed targets have colors far from those expected from simple stellar population models, which is likely due to strong emission lines in their spectra (e.g., Tran et al.~2003; Gallagher~et~al.~2010; Konstantopoulos et~al.~2010). Circles show sources that have strong emission lines in their spectra.  In particular, the F606W (V-band) filter is quite broad and will contain contamination from many of the strong emission lines in the optical, namely H$\beta$, [OIII], H$\alpha$, [NII], and [SII].  

While spectroscopy is a powerful way to break the age/extinction degeneracy that affects photometric measurements of young clusters, the low spatial resolution of ground based observations gives rise, in some cases, to multiple sources contributing to the observed spectrum. Fig.~\ref{fig:images} shows close-up color images of all sources. Many of the targets, especially the ones associated with recent/active sites of star formation, are not single clusters, but cluster complexes.  Hence, we will refer to all sources as clusters/complexes to denote their status, although we note that some of the older sources (e.g. T118, T122, and T123) do appear to be single clusters.

\section{Derivation of Cluster Properties}
\label{sec:properties}

We have developed a sequence of routines to estimate the age and metallicity of young stellar clusters from optical spectra.  Details of the method are given in T07b and we briefly outline the technique here.  Targets with spectra that are dominated by absorption (stellar photospheres) or emission (surrounding ionized gas) are treated separately. 

For absorption line-dominated spectra, we first construct a template for each cluster using the Penalized Pixel-Fitting (pPXF) method (Cappellari \& Emsellem 2004).  The method is based on the Bounded-Variables Least-Squares algorithm and constructs a template based on a combination of simple stellar population (SSP) models, and uses a penalized maximum likelihood formalism.  It has the advantage of being robust even when the data have a low signal-to-noise ratio (S/N).   The pPXF method, as realized in an IDL routine, takes in SSP model spectra of different ages/metallicities and weights them in order to create a best fit stellar template spectrum.  For the SSP models we chose the Gonz\'alez-Delgado et al.~(2005) models (of 1/5, 2/5, 1, and 2.5 times solar; Salpeter stellar IMF; and ages between 4~Myr and 15~Gyr) due to their high spectral and temporal resolution.

\begin{figure}
     \plotone{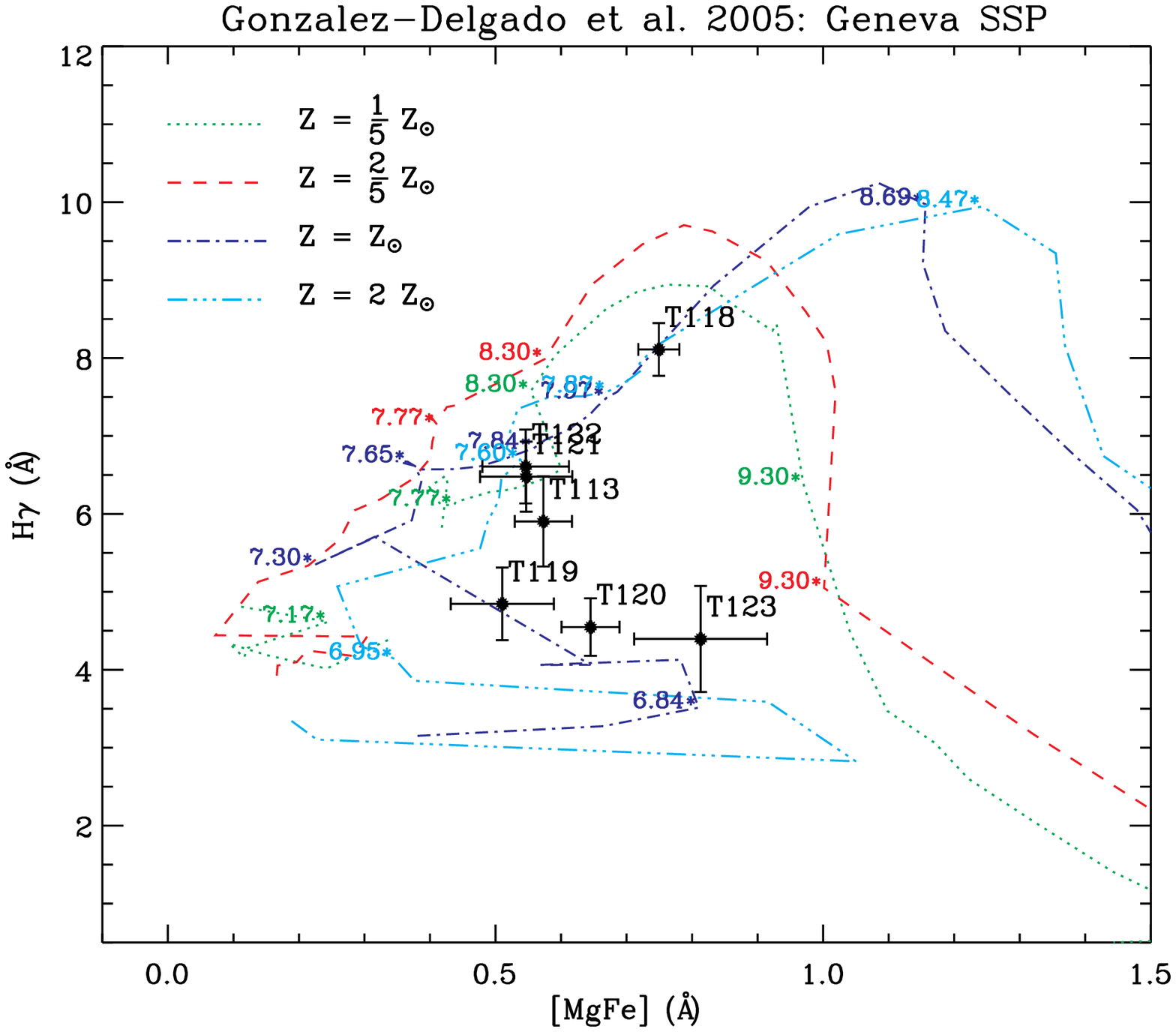}
      \caption{Determination of ages and metallicities of the
     clusters.  H$\gamma$ vs. [MgFe] from the Gonzalez-Delgado, 2005 (GD05) SSP models for four different metallicities are shown.
     The positions and errors ($1\sigma$) are shown. }
 \label{fig:index}
\end{figure}

After, we measure the strengths of a series of metal and Balmer lines (line indices after Schweizer~\&~Seitzer~1998),  which are given in Table~\ref{table:indices}. These are then compared to model SSP spectra (Gonz\'alez-Delgado et al.~2005) whose spectral indices are measured in an identical way.  In Fig.~\ref{fig:index} we show the resulting H$\gamma$ and [MgFe] indices for the clusters, along with the predictions from SSP models. 

For the emission line-dominated targets we assign ages less than 7~Myr and estimate the extinction from the ratio of H$\gamma$ to H$\beta$, or H$\beta$ to H$\alpha$, depending on spectral coverage.  The metallicites are then estimated using the R$_{23}$ and O$_{32}$ methods, as described in Kobulnicky \& Kewley (2004)\footnote{
	$\log R_{23} = \frac{I_{\textup{\scriptsize [O\,{\sc ii}]}\,\lambda3727} + I_{\textup{\scriptsize [O\,{\sc iii}]}\,\lambda4959} + I_{\textup{\scriptsize [O\,{\sc iii}]}\,\lambda5007}}{I_{\textup{\scriptsize H}\beta}}$, \\
	$\log O_{32} = \log(\frac{I_{\textup{\scriptsize [O\,{\sc iii}]}\,\lambda4959} + I_{\textup{\scriptsize [O\,{\sc iii}]}\,\lambda5007}}{I_{\textup{\scriptsize [O\,{\sc ii}]}\,\lambda3727}})$
	}. These methods employ equivalent widths, rather than line fluxes, and as such, are independent of extinction. Line widths are shown in Table~\ref{table:indices2}. We implicitly assume that the clusters/complexes are on the ``upper-branch" of the R$_{23}$ vs. oxygen abundance relation, which is consistent with the metallicities found for the absorption line clusters that is independent of any such assumptions.
The details of this process are given in T07a, T07b, and B09.  

The derived ages and metallicities of all clusters/complexes are given in Table~\ref{table:properties}.  

\subsection{Velocities}
In the case of absorption-line clusters, we used the IRAF task
{\em rvsao.xcsao} for the determination of the redshift from the
individual spectra, using radial-velocity standard stars of three different types: A, O, and B (HD~100953, HD~126248, and HD~133955) observed at the same resolution as the clusters.
The three template stars were employed to reduce the
systematic errors introduced by the effect of template mismatch when
computing the redshift through cross-correlation.

For the emission-line clusters, velocities were measured from the observed emission lines using the IRAF task {\em rvsao.emsao}.  In both cases the velocities were corrected to the heliocentric system, and are presented in Table~\ref{table:properties2}.

\subsection{Extinctions and masses}

In our previous study of the Antenn\ae\ galaxies, we used photometry in order to estimate the extinction to each cluster.  By comparing the broad-band photometry (UBVI) to SSP models we could derive the extinction as the age was known from spectroscopy.  In the present case however, this is not possible, as the BVI colors of clusters are highly degenerate for ages between 10 and a few hundred Myr (see Fig.~\ref{fig:cc}).  However, since the clusters appear to follow the SSP model track well, it appears that the overall extinction for each cluster in our sample is rather low (A$_{\rm V} < 1$). This is a likely selection effect, as highly extinguished knots would either be beyond the diagnostic reach of our method, or simply not targetted in favour of including brighter targets.  

The mass of each cluster/complex was estimated through a comparison of the observed brightness to that of an SSP model at the spectroscopically derived age. Photometry was carried out in apertures matching the width of the spectroscopic slits. We have only corrected the photometry for Galactic extinction, hence the estimated masses, given in Table~\ref{table:properties2}, are likely to be lower-limits.

\subsection{Contamination}

Many of the targets included in the current survey were already studied in earlier \hst\ imaging by G01.  In Tables~\ref{table:comparison1}~and~\ref{table:comparison2}  we show the list of targets that overlap with G01 and their classification.  We confirm the membership of many G01 candidate clusters, while some of the redder sources are foreground stars.  One object of particular interest is T118, which was noted by G01 as a potentially massive cluster. This cluster is confirmed as a member, and we will discuss it in detail in \S~\ref{sec:t118}.

In order to facilitate photometric studies of the cluster population in SQ (e.\,g. F11; Konstantopoulos~et~al.~in~prep) we show the measured photometric properties of all our targets in Fig.~\ref{fig:cc}.  Some of the G01 and F11 
candidates were found to be foreground stars, indicating the need to take into account contamination when selecting target clusters, even for galaxies $\sim$20 degrees from the Galactic plane. Restricting the analysis to targets with F606W$ - $F814W, i.\,e. $V-I < 0.7$ appears to effectively result in a clean sample. Additionally, F11 have been able to show a clear { concentration} of red (old) globular clusters ($V-I > 0.7$) around a number of SQ galaxies, showing that while individual candidates may be uncertain, the population can still be studied. However, we find little or no contamination in our sample for ages $< 1$~Gyr.

\section{Results}
\label{sec:results}

As seen in Table~\ref{table:properties} the majority of the clusters in our sample are young, with ages less than $\sim20$~Myr.  This likely represents an observational bias: due to the distance to the system we could only target the brightest clusters of the system, which are preferentially young (e.g. Bastian~2008; Larsen~2009; Gieles~2010).  The exception, T118, will be discussed in detail in \S~\ref{sec:t118}.  While spectroscopic age determinations are clearly more robust than photometry alone, the small sample sizes and strong observational selection effects limit the applicability of this method to constrain the cluster formation history of distant systems or to study cluster disruption mechanisms.

Based on deep WFC3 B, V, and I band imaging, F11 have been able to find differences in ages between cluster populations located throughout SQ.  While the age and extinction of simple stellar populations are largely degenerate when only using B, V, and I, F11 found clear evidence for differing formation histories of cluster populations in different parts of SQ. They interpreted the age distributions in the context of the interaction histories of the galaxies, guided by dynamical models of the system. Here, we are able to directly test their estimated ages (and age spreads) through spectroscopy, which does not suffer from an age/extinction degeneracy.  The targets covered in this work are found in the regions identified by F11 (see Fig.~\ref{fig:image_label}) as 1) the Northern Starburst Region (NSBR), 2) NGC 7318A/B, 3) the Southern Debris Region (SDR), and 4) the Young Tail. Here we analyze each region separately and compare our derived ages with those of F11.

In (nearly) all regions studied, F11 found a population of old globular clusters, that are presumably associated with the haloes of the SQ galaxies. Given their old ages, these clusters are much fainter ($V > 24$) than young clusters, and hence were not targeted in our program, which can only reach V$\sim 21~$mag.

\subsection{The Northern Starburst Region (NSBR)} 

This region, located approximately 30" north of NGC 7318B and 45" west of NGC 7319, is the site (at least in projection) where the two tidal tails to the north of NGC 7318A/B overlap, and the region has a significant amount of ongoing star formation (F11).  Dynamical models have suggested that the two tidal tails are currently physically interacting, triggering the ongoing burst of star formation (Renaud et al.~2010) which is supported by H{\sc i} velocity maps presented by Williams et al.~(2002).  F11 find evidence for many young SCCs ($<10$~Myr) in the region as well as some clusters with colors consistent with either a reddened (A$_{V}  = 0.5$~mag) young population, or a non-reddened older ($>100$~Myr) population.

We have five confirmed clusters/complexes in the region, T110a-e, all of which display strong emission lines, indicating that they are quite young ($<7$~Myr). { The ratios of their emission lines are indicative of H{\sc ii} region ionization, rather than arising due to shocks (see Section~\ref{sec:e-spec}).} Hence, we confirm the ongoing starburst nature of the region put forward by F11, but cannot probe the suggested older clusters, since they would be below our brightness limits to obtain spectroscopy.

The velocities of the complexes in this region agree well with the surrounding H{\sc i} gas (Table~\ref{table:properties2}), showing a gradient of increasing velocity away from NGC 7318A.  Additionally, we can see which young region is associated with which H{\sc i} component (i.e. which tail); T110a-d have velocities consistent with the low velocity tail (associated with NGC~7318A), while T110e has a velocity $>600$~km/s different from the nearest young region (T110d) and appears to be associated with the tail originating from NGC 7318B.  The high relative velocities associated with different complexes within this region prove its transient nature.

\subsection{NGC 7318A/B}

NGC~7318 A and B are the two galaxies within SQ that are currently strongly interacting. Because of their proximity we analyze the population of SSCs around these galaxies together, as did F11. Their photometric analysis found evidence for ongoing cluster formation within this region for the past $250$~Myr and a distinct lack of clusters with ages between $\sim400$~Myr and $\sim2$~Gyr.  NGC~7318A is an { early type} galaxy (found in a GMOS spectrum obtained by Konstantopoulos et al., in prep) while NGC~7318B is a spiral galaxy thought to be falling into SQ with a high velocity for the first time (Moles et al.~1997).

Our sample contains nine clusters/complexes within these galaxies and the associated debris.  T119, T120, T121 and T123 are associated with the tidal tail/spiral arm emanating from NGC 7318B, while T111a-d and T112 all lie to the south of NGC~7318A, with its associated debris.  All clusters in this region are young ($<20$~Myr) and the clusters associated with the tidal tail of NGC~7318B follow a clear velocity gradient, as expected from numerical simulations (Renaud et al.~2010) and observed gas distributions (e.g., Williams et al.~2002, Plana et al.~1999). 

Clusters/complexes T111a-d and T112 do not have similar velocities to NGC~7318A, but instead are consistent with NGC~7318B.  This implies that they formed from debris stripped from this spiral galaxy, and form a continuation of the spiral arm and tidal tail originating from NGC~7318B (as also supported by Renaud et~al.~2010).

\subsection{Southern Debris Region (SDR)}

The Southern Debris Region (SDR) lies $\sim45$" to the south-west of NGC~7318A and hosts a number of extended low surface brightness sources.  F11 find a large population of old globular clusters in the region, which are most likely associated with the nearby (in projection) elliptical galaxy NGC~7317.  They also find young SSCs with colors suggesting ages between 50 and 400~Myr, and a handful of SSCs with colors suggesting the presense of emission lines.  The four confirmed clusters/complexes in our sample that lie within this region (T113-116a,b) are all quite young, with ages $\lesssim7$~Myr.  This raises the possibility that the clusters in the F11 sample of this region are all/mostly young and are affected by extinction. This is consistent with F11, who find a large population of `nebular' clusters in the region, i.\,e. ones affected by nebular emission covered in the F606W filter. 

The velocities of the clusters/complexes in the SDR are all consistent with H{\sc i} gas in the region, as well as the shock front of NGC~7318B and NGC~7318B itself.  These complexes are also consistent with the velocity gradient of T111a,b,c and T112, suggesting that this feature may be a continuation of the debris extending from NGC~7318B.

We note that T114 (along with other clusters/complexes in our sample) have relatively high metallicities, being super-solar.  This is not unexpected, as the debris from which these clusters are forming has come directly from a spiral galaxy.  Rupke et al.~(2010) have also found super-solar metallicity emission line regions outside galaxies in their sample of interacting/merging systems.

\subsection{Young Tail}

The Young Tail is a tidal feature that is thought to have originated during an interaction between NGC~7318A and NGC 7319 approximately $200$~Myr ago (Renaud et al.~2010). %
 F11 find a large number of SSCs associated with the tail, with a remarkably small dispersion in their colors ($B-V = 0.22\pm0.07$ and $V-I = 0.39\pm0.12$~mag, where the errors are the standard deviations), suggesting an age of $\sim200$~Myr.  They found only four SSC candidates outside this small locus of points, { which may not be old but young, highly extinguished clusters}.  

Four of our confirmed sources lie within the Young Tail (T117, T118, T122 and T124), however only T118 lies within the WFC3 field of view used by F11.  The derived age of T118 is $\sim125$~Myr, is in good agreement with the mean age estimate of F11 ($\sim200$~Myr) and G01 ($\sim150$~Myr)\footnote{The different mean ages for the region arise from different coverage in the pointings used by G01 and F11.}.
T118 will be discussed in more detail in \S~\ref{sec:t118}.  Concerning the other three sources in the Young Tail: 
T117 has strong emission lines indicating an age $<7$~Myr; 
T122 is absorption line-dominated indicating an age 50~Myr; 
and T124 (the tidal dwarf galaxy candidate) has an age of $\sim7$~Myr and will be discussed in detail in \S~\ref{sec:tdg}. Hence, at least in the outer regions of the tail, we find that star formation is ongoing, suggesting that the tidal debris of galaxy mergers can have extended periods ($>125$~Myr) of star formation.  Such an extended star formation history has also been found for three clusters in the tidal tail of NGC~3256 (T07a).

\subsection{Formation of a massive cluster within the tidal debris}
\label{sec:t118}

One cluster in our sample, T118, stands out as particularly interesting. It is located in the easternmost tidal debris, and its position is indicated in Fig.~\ref{fig:image}.  This source was singled out by G01 as its brightness rivals that of clusters normally found only in merging and/or starburst galaxies.

From its spectrum we derived an age of $125^{+75}_{-50}$~Myr that, given its brightness (M$_{\rm V} = -13.2$; G01), corresponds to a mass of  {$\sim2.2\times10^6~$\msun}.  Its age and compact structure show that it is likely to evolve into a bound and long-lived cluster.  Since the cluster is located far from the main bodies of any of the galaxies in this system, it is not likely to lose a significant amount of mass from tidal stripping (e.g. Baumgardt \& Makino~2003; Lamers et al.~2005).  Hence, T118 is expected to survive for a long time, and is likely to become an intergalactic star cluster, like that seen in many galaxy groups (e.g. West et al.~1995).

Given that T118 has colors similar to many other objects along the Young Tail, it is likely to simply be the most massive member of a large, but undetected, underlying population.  We can compare the properties of T118 with other massive SSC candidates in tidal structures from Mullan et al.~(2011).  From their sample, we see that three galaxies, NGC~2782, AM~1054-325 and NGC~6872, host SSC candidates with masses near $1\times10^6\msun$ in their tidal debris.  Mullan et al.~(2011) conclude that the cluster populations in tidal debris follow the same statistical relations as populations in normal galaxies (e.\,g. number vs. M$_{V}^{\rm brightest}$; Larsen~2002, Whitmore~2003; Gieles et al. 2006), suggesting that star/cluster formation does not operate significantly differently outside galaxies. 

Based on the WFC3 V-band image, T118 appears to be slightly resolved relative to stars in the field-of-view.  We used {\it ISHAPE} (Larsen~1999) to determine the effective radius of the cluster, adopting a Moffat profile with index 1.5 (typical for young massive clusters - Larsen~2004).  With an effective radius of $\sim12$~pc, T118 appears to be much larger than typical SSCs ($\sim3$~pc - Larsen 2004; Barmby et al.~2006; Scheepmaker et al.~2007; Portegies Zwart et al.~2010).  However, previous studies of the sizes of clusters within tidal debris of the merging galaxy NGC~3256 have found that these clusters are much larger ($10-20$~pc) than normal (T07a), hence T118 is consistent with the { cluster formation effect} suggested by T07a.  Whether this size difference is due to the lack of tidal stripping of the outer parts of the cluster or is due the formation of the cluster is cannot be determined with the present data.

Several of the attributes of this object qualify it as an ultra-compact dwarf galaxy (UCD), namely its high mass and extended nature (e.g., Mieske et al. 2008). However, its young age and the fact that it has similar colors to surrounding, lower luminosity, objects argues that it is a stellar cluster, and simply the most massive in a continuous distribution (e.g., Maraston et al. 2004).  

\subsection{A Tidal Dwarf Galaxy candidate} 
\label{sec:tdg}

\begin{figure}
     \includegraphics[width=0.5\linewidth]{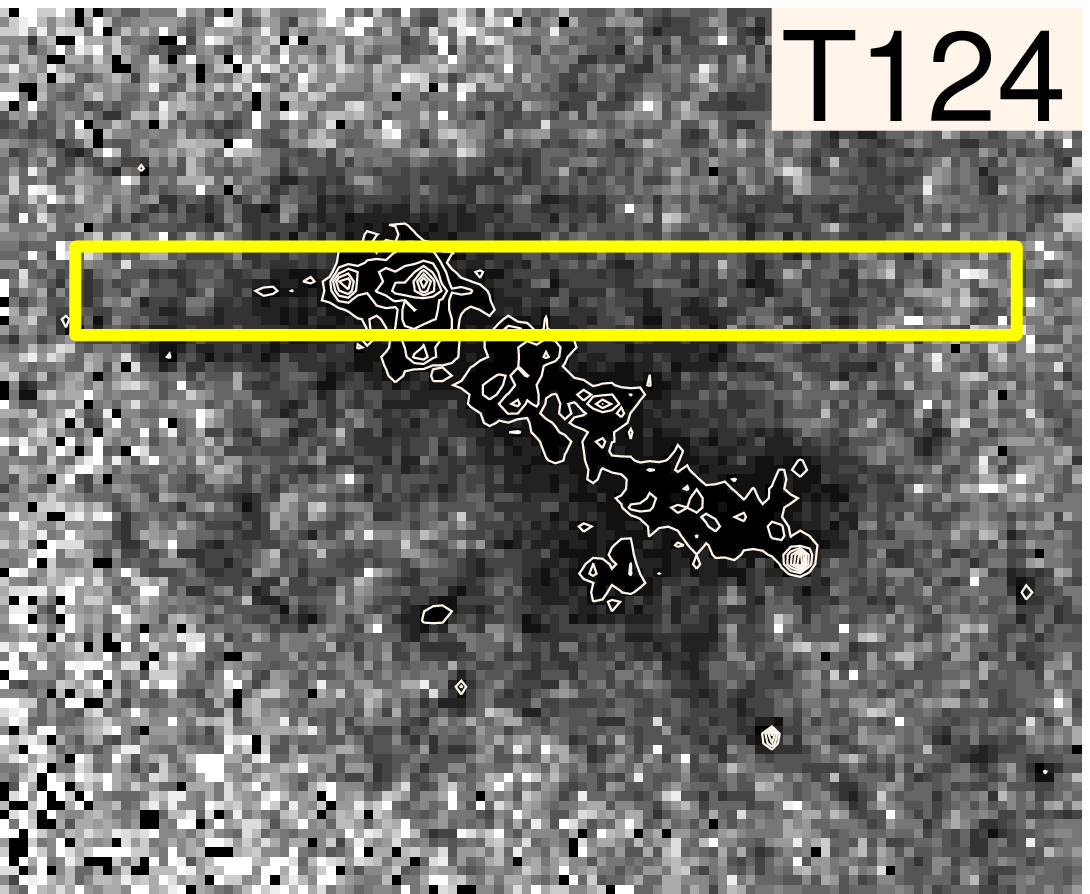}
     \includegraphics[width=0.5\linewidth]{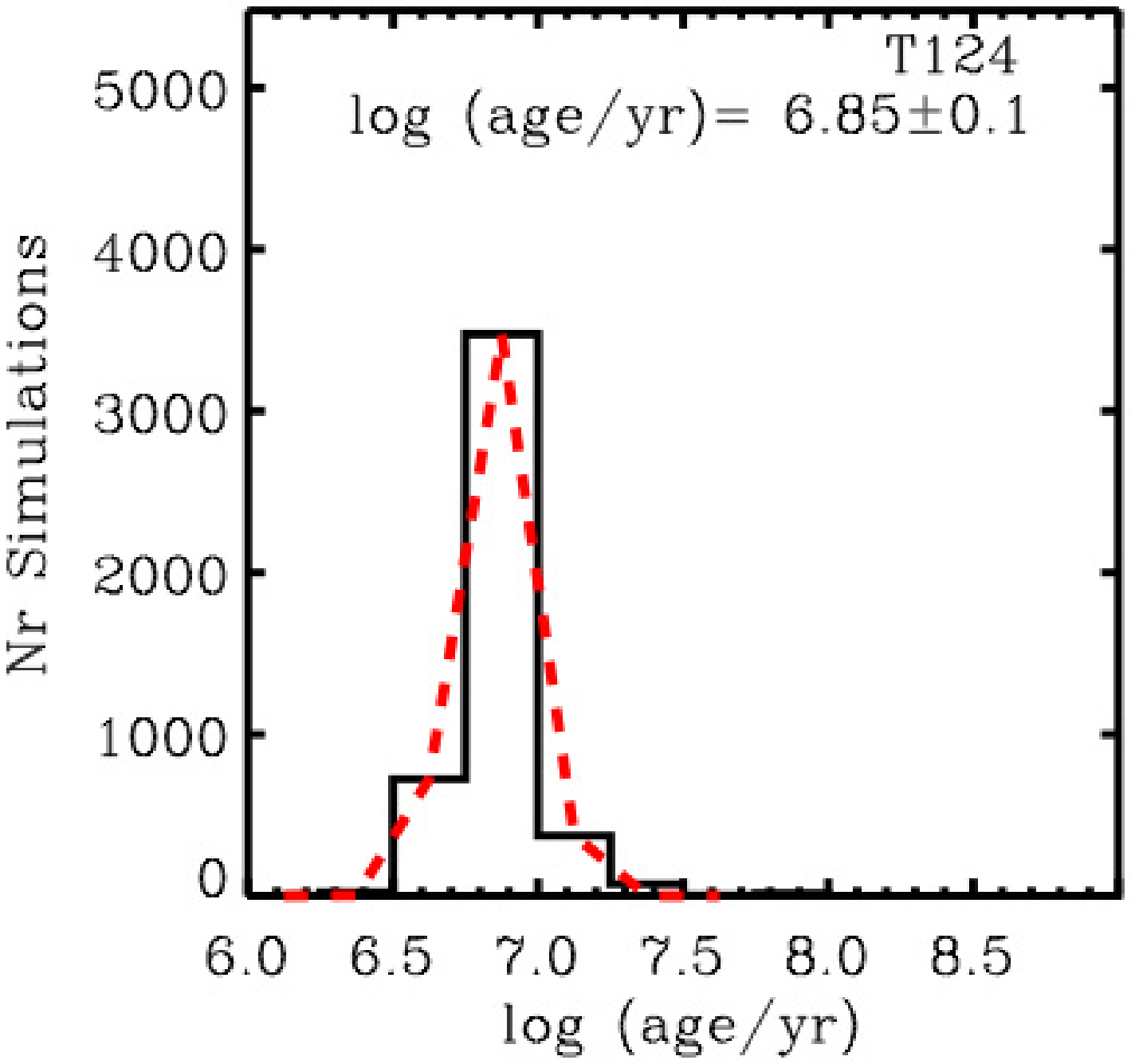}
      \caption{{ Left panel:}  A WFPC2 V-band image of the Tidal Dwarf Galaxy Candidate at the end of the Young Tail.  Contours show the flux levels relative to the maximum intensity of the image at 10\% intervals.  The image is 115 by 100 WF~pixels, corresponding to 11.5 by 10 arcseconds or 4.7 by 4.2 kpc at the adopted distance.  { Right panel:}  The distribution of derived ages for the TDG based on monte carlo sampling of the observational errors (see T07b for a full description).  The TDG is clearly dominated (in luminosity at least) by a young population.}
 	\label{fig:tdg}
   \end{figure}

In addition to the 21 clusters/complexes discussed above, we have also obtained a spectrum of the Tidal Dwarf Galaxy (TDG) candidate at the end of the Young Tail.  A WFPC2 V-band image of this low surface brightness object is shown in Fig.~\ref{fig:tdg}, along with the slit position.   Although this is a resolved object, and likely has undergone an extended star formation history, the obtained spectrum does not have the necessary S/N to attempt a multi-component fit of age, extinction and metallicity.  Instead, we have adopted the same technique as for the clusters, and assume that it can be described by a single stellar population.  This is justified to some extent by the lack of a strong color gradient along the TDG, as measured on the \hst\ images. 

The spectrum of T124 does not show prominent emission lines, but a fit to the spectral continuum shape reveals a young age, $\sim7$~Myr, and a metallicity consistent with other clusters in the Young Tail ($\sim1.4~Z_{\odot} \pm 0.4$). 
{ The above interpretation assumes a single age for the entire clump, given the aforementioned lack of S/N (and resolution) required to perform a multi-component fit}. 
Using the same method as above, we have estimate the mass contained in 1) a 10-pixel aperture of the end clump where our spectrum was obtained and 2) a 35 pixel aperture which covers the full extent of the TDG.  We obtain masses of $\sim3.3$ and $\sim10.3 \times 10^5$\msun, respectively. { These values both represent lower limits on the mass of this clump, given the temporally increasing mass-to-light ratio of simple stellar populations. In that way, the same flux would yield a higher mass if the SSP was assumed to be older. An order of magnitude increase in the age estimate would give rise to a similar increase in the mass calculation.}

{ Assuming this clump is a young SSP, certain characteristics can be derived}. While this is clearly a massive system, its low surface brightness (an average of $\sim22.8$ mag arcsec$^{-2}$ in the V-band within the 35 pixel aperture) and young age means that it will fade by $\sim2$~mags by the time it reaches an age of 50~Myr, assuming that it does not expand further.  Hence, from an observational point of view, this is a transient feature, although it may remain a long lived feature.

\section{Conclusions}
\label{sec:conclusions}

We have presented optical spectroscopy of 21 young massive clusters/cluster complexes and one tidal dwarf galaxy (TDG) candidate in the interacting compact galaxy group, Stephan's Quintet.  From the spectroscopy and archival HST imaging, we have estimated the age, metallicity, mass and extinction of each of the clusters/complexes.  Overall, our derived ages agree well with Fedotov et al.~(2011) who estimated the ages of sub-populations of clusters in various regions of the Stephan's Quintet using B, V and I WFC3 imaging.  Additionally, we confirm the sequence of merger events that have spawned various sub-populations as postulated by Renaud et al.~(2010) using numerical simulations and bolstered by the Fedotov et al~(2011) photometry.

One cluster in our survey, T118 (in the young tidal tail region), is a particularly interesting object, given its relatively large age ($\sim125$~Myr) and high mass ($\sim2 \times 10^6$~\msun).  This object is likely to survive for an extended period, becoming an intergalactic star cluster, similar to that seen in other galaxy groups (e.g. West et al.~1995).  T118 also appears extended in the images, with an estimated effective radius of $\sim12$~pc, which is much larger than typical young massive clusters (R$_{\rm eff} \sim 3-4$~pc).  Extended (R$_{\rm eff} > 10$~pc) massive clusters have also been observed in the tidal debris of NGC~3256 (T07a), suggesting a general trend that the environment of tidal debris is particularly suitable for forming or nurturing extended clusters.  The similar colors of many clusters along the young tidal tail as T118, suggest that T118 is simply the most massive cluster in a continuous population.  Clusters in the Young Tail have ages ranging from active sites of star-formation to $\gtrsim125$~Myr, suggesting that star/cluster formation within the tidal debris can extend for significant periods of time.

We have also obtained a spectrum of the TDG at the tip of the Young Tidal Tail and find a very young age, $\sim7$~Myr.  Unfortunately, due to the low S/N of the spectrum we were not able to undertake a multi-component fit to the age derivation.  Based on this single age fit, and the flux estimated for the full TDG, we estimate a mass of $\sim10^6$\msun.  However, the low surface brightness of the region and young age, means that it will be an observationally transient feature as it is expected to fade beneath observational detection limits with the next 10 to 100 Myr.  However, it may remain a physically long-lived and coherent feature due to the weak tidal fields in the far tidal debris of the Stephan's Quintent.


In an upcoming paper, Konstantopoulos~et~al. will study the physics of the shocked intra-group medium in SQ through newly obtained GMOS spectroscopy. 

\begin{acknowledgements}
ISK acknowledges the support of Gemini Observatory for a research assistantship, in the bounds of which part of this work was undertaken. ISK and JCC acknowledge support by the National Science Foundation under award 0908984. NB was supported by the DFG cluster of excellence 'Origin and Structure of the Universe'.  KF and SCG thank the National Science and Engineering Research Council of Canada and the Ontario Early Researcher Award Program for support. Based on observations obtained at the Gemini Observatory, which is operated by the Association of Universities for Research in Astronomy, Inc., under a cooperative agreement with the NSF on behalf of the Gemini partnership: the National Science Foundation (United States), the Science and Technology Facilities Council (United Kingdom), the National Research Council (Canada), CONICYT (Chile), the Australian Research Council (Australia), Minist\'erio da Ci\'encia e Tecnologia (Brazil) and Ministerio de Ci\'encia, Tecnolog\'ia e Innovaci\'on Productiva  (Argentina).

\end{acknowledgements}

\begin{deluxetable}{lcccccccccc}
\def\psn{\phs\phn}
\def\pnn{\phn\phn}
\tablecolumns{9}
\tablewidth{-10pt}
\tablecaption{Derived properties of star clusters in Stephan's Quintet}
\tablehead{
\colhead{ID} & \colhead{A/E\tablenotemark{a}} &  \colhead{RA\tablenotemark{b}} &  \colhead{DEC\tablenotemark{b}}  & \colhead{F438W\tablenotemark{c}} & \colhead{F606W\tablenotemark{c}}& \colhead{F814W\tablenotemark{c}}& \colhead{A$_V$} & \colhead{Z}& \colhead{Log(age)}\\
\colhead{ } & \colhead{ } &   \colhead{(sec) } & \colhead{(arsec) } & \colhead{(mag)} & \colhead{(mag)} &  \colhead{(mag)} & \colhead{(mag)} & \colhead{(\zo) }& \colhead{(year)}  } 
\startdata
T110a	&1    &22:35:57.61&+33:59:01.06 & 23.27 & 22.89 & 22.86 &-     & -           & $<$6.8      \\  
T110b	&1    &22:35:57.95&+33:58:58.86 & 23.49 & 22.68 & 23.18 &-     & -           & $<$6.8      \\ 
T110c	&1    &22:35:58.27&+33:58:55.46 & 23.41 & 23.17 & 23.09 &-     & -           & $<$6.8      \\ 
T110d	&1    &22:35:58.47&+33:58:53.36 & 22.18 & 22.08 & 21.89 &-     & -           & $<$6.8      \\ 
T110e	&1    &22:35:58.97&+33:58:49.86 & 21.63 & 20.72 & 20.36 &-     & -           & $<$6.8      \\ 
T111a	&1    &22:35:56.74&+33:57:38.56 & 22.23 & 22.04 & 22.04 & 0.4 & 1.1$\pm$0.2 & $<$6.8      \\ 
T111b	&1    &22:35:56.97&+33:57:37.63 & 23.04 & 22.86 & 22.77 & 0.4 & 1.0$\pm$0.2 & $<$6.8      \\ 
T111c	&1    &22:35:57.22&+33:57:36.43 & 23.25 & 23.13 & 23.05 & 0.6 & 0.7$\pm$0.2 & $<$6.8      \\ 
T112	&1    &22:35:55.67&+33:57:43.56 & 22.28 & 21.97 & 21.81 &-     & 1.1$\pm$0.3 & $<$6.8 \\ 
T113	&0    &22:35:55.48&+33:57:10.56 & 22.65 & 22.44 & 21.90 &-     & 1.0$\pm$0.4 & 7.7$\pm$0.1 \\ 
T114	&1    &22:35:55.07&+33:57:10.56 & 23.76 & 23.32 & 23.53 &-     & 1.5$\pm$0.2 & $<$6.8 \\ 
T115	&1    &22:35:54.98&+33:57:13.06 & 23.92 & 23.83 & 23.66 &-     & 0.8$\pm$0.2 & $<$6.8      \\ 
T116a\tablenotemark{d}%
		&1    &22:35:54.19&+33:57:16.06 & 22.55 & 22.36 & 22.12 &-     & -           & $<$6.8      \\ 
T116b	&1    &22:35:54.29&+33:57:15.46 & 22.45 & 22.34 & 21.99 &-     & 0.8$\pm$0.2 & $<$6.8      \\ 
T117	&1    &22:36:10.28&+33:57:20.96 & 24.00 & 22.09 & 21.62 &2.0& 1.0$\pm$0.2 & $<$6.8      \\ 
T118	&0    &22:36:08.77&+33:57:20.96 & 21.94 & 21.70 & 21.41 & 0.2 & 0.6$\pm$0.2 & 8.1$\pm$0.2 \\  
T119	&0    &22:36:00.41&+33:58:03.56 & 21.92 & 21.79 & 21.41 & 0.4 & 0.7$\pm$0.3 & 7.1$\pm$0.1 \\  
T120	&0    &22:36:00.03&+33:57:46.46 & 22.10 & 21.99 & 21.55 & 0.0 & 1.0$\pm$0.4 & 7.1$\pm$0.1 \\  
T121	&0    &22:35:59.77&+33:57:38.96 & 22.01 & 22.06 & 21.94 & 0.5 & 1.5$\pm$0.3 & 7.6$\pm$0.1 \\  
T122	&0    &22:36:10.25&+33:57:23.26 & 24.39 & 24.07 & 23.68 & 0.0 & 1.0$\pm$0.4& 7.7$\pm$0.1      \\  
T123	&0    &22:35:59.31&+33:58:15.06 & 22.68 & 22.50 & 22.15 & 0.4 & 0.7$\pm$0.2 & 7.0$\pm$0.1 \\  
T124\tablenotemark{e} &0  &22:36:19.35&+33:57:43.60 &  19.90\tablenotemark{f} & 19.70\tablenotemark{g} & 19.00\tablenotemark{f}  & - &  1.4$\pm$0.4 & 6.8$\pm$0.1 \\ 
T124\tablenotemark{g} &0  &22:36:19.35&+33:57:43.60 &  21.50\tablenotemark{f} & 20.95\tablenotemark{f} & 20.88\tablenotemark{f}  & - &  1.4$\pm$0.4 & 6.8$\pm$0.1 \\ 
\enddata
\tablenotetext{a}{\,0=absorption, 1=emission}
\tablenotetext{b}{\,RA,DEC (J2000).}
\tablenotetext{c}{\,The magnitudes have been only corrected for Galactic extinction.}
\tablenotetext{d}{\,We were only able to measure the velocity of cluster T116a, and not other spectroscopic properties. }
\tablenotetext{e}{\,The properties of the TDG candidate, derived within an aperture of 35~pixels. 
The age is measured for the part of the clump covered in our spectroscopic slit, not the entire complex.}
\tablenotetext{f}{\,WFPC2 images used, magnitudes given in the Cousins-Johnson B, V, and I-bands.}
\tablenotetext{g}{\,The properties of the TDG, this time confined to the region covered by our spectroscopic slit, as shown in Fig.~\ref{fig:tdg}.}

\label{table:properties}
\end{deluxetable}


\begin{deluxetable}{lccccc}

\tablecaption{Spectrophotometry of all emission-line clusters.}
\tablewidth{-10pt}

\tablehead{\colhead{ID}								&	 
			\colhead{$F$(H$\beta$)}					&	 
			\colhead{$F$([OIII]$_{\lambda4959+5007}$)}   &	 
			\colhead{$F$(H$\alpha$)}				&	 
			\colhead{$F$([NII]$_{\lambda6548+6584}$)}	&	 
			\colhead{$F$([SII]$_{\lambda6713+6731}$)}	\\	 
			\colhead{}  							&							
			\multicolumn{4}{c}{\tlabel}}
\startdata
 T110a &  0.25 &  0.31 &  1.07 &  0.35 & 0.19 \\
 T110b &  0.15 &  0.24 &  0.55 &  0.15 &0.11 \\
 T110c &  0.09 &  0.11 &  0.40 &  0.13 &0.10 \\
 T110d &  0.19 &  0.32 &  0.85 &  0.13 & 0.16 \\
 T110e &  0.64 &  1.65 &  3.51 &  0.40 & 0.55 \\
 T111a &  0.18 &  0.17 &   $-$ &   $-$ &$-$ \\
 T111b &  0.06 &  0.06 &   $-$ &   $-$ &$-$ \\
 T111c &  0.08 &  0.08 &   $-$ &   $-$ &$-$ \\
 T112  &  0.47 &  0.90 &   $-$ &   $-$ &$-$ \\
 T114  &  0.09 &  0.22 &   $-$ &   $-$ &$-$\\
 T115  &  0.03 &  0.03 &   $-$ &   $-$ &$-$ \\
 T116b\tablenotemark{a} &  0.08 &  0.17 &   $-$ &   $-$ &$-$ \\
 T117  &  0.16 &  0.37 &   $-$ &   $-$ &$-$ \\
\enddata

\tablenotetext{a}{T116a is not listed, as no fluxes could be measured. }
\label{table:specphot}
\end{deluxetable}

\begin{deluxetable}{lccccccccc}
\tablecaption{Kinematics and Masses of the clusters.}
\tablewidth{-10pt}

\tablehead{
\colhead{ID} & \colhead{D\tablenotemark{a}}  & \colhead{cz(CO)\tablenotemark{b}}& \colhead{cz(\hi)\tablenotemark{c}}  & \colhead{cz$_{hel}$} & \colhead{$\delta_{cz}$}\tablenotemark{d}  &  \colhead{Mass}  & \colhead{Region\tablenotemark{e}}\\ 
\colhead{} & \colhead{} & \colhead{(km/s)} &\colhead{(km/s)} & \colhead{(km/s)} & \colhead{(km/s)} & \colhead{$10^5$\msun} & \colhead{}} 

\startdata
T110a& 1  &-     	&5960&    5909.2$\pm$25.3 &+50.8	& 0.31&  NSBR&\\
T110b& 1  &-  		&5960&    5937.7$\pm$15.1 &+22.3 & 0.38 &  NSBR&\\
T110c& 1  &-  		&6046&    6045.4$\pm$28.2 &+0.5 	& 0.24&  NSBR&\\
T110d& 1  &-  		&5960&    5971.0$\pm$39.2 &-11  	& 0.66&  NSBR&\\
T110e& 1  &-  		&6670&    6661.0$\pm$27.2 &+9  	& 2.33&  NSBR&\\
T111a& 1  &-  		&5710&    5700.2$\pm$27.4 &+9.2  	& 0.69&  NGC~7318A/B&\\
T111b& 1  &-  		&5661&    5660.0$\pm$14.1 &+1  	& 0.32&  NGC~7318A/B&\\
T111c& 1  &-  		&5710&    5717.1$\pm$10.5 &+6.9  	& 0.25&  NGC~7318A/B&\\
T112& 1    &-  		&5740&    5743.2$\pm$22.3 &-3.2  	& 1.45&  NGC~7318A/B&\\
T113& 0    &-  		&5740&    5750.0$\pm$14.6 &-10  	& 0.94&  SDR&\\
T114& 1    &-  		&5750&    5700.7$\pm$10.9 &+49.3 & 0.42&  SDR&\\
T115& 1    &-  		&5750&    5756.0$\pm$43.7 & -6 	& 0.13&  SDR&\\
T116a& 1  &- 		&5770&    5777.8$\pm$58.0 &-7.8  	& 0.51&  SDR&\\
T116b& 1  &-  		&5770&    5764.0$\pm$15.7 &+6  	& 0.52&  SDR&\\
T117&  1    &6614     &6604 &   6603.8$\pm$12.2 &+9.2/+0.2& 0.66 &  Young Tail&\\
T118&  0    &-  		&6610&    6635.7$\pm$19.3 &-19.3  & 22.58&  Young Tail &\\
T119&  0    &-  		&-	  &    5825.7$\pm$24.1 &-   	& 6.96&  NGC~7318A/B&\\
T120&  0    &-  		&- 	  &    5695.1$\pm$19.5 &-		& 3.07&  NGC~7318A/B&\\
T121&  0    &-  		&-	  &    5701.6$\pm$45.2 &- 		&1.34 & NGC~7318A/B &\\
T122&  0    &6614     &6604&    6640.3$\pm$19.6 &-26.3/-36.3& 0.21&  Young Tail&\\
T123&  0    &-  		&- 	  &    6021.4$\pm$33.8 &-  	& 1.92&  NGC~7318A/B&\\
T124\tablenotemark{e}&  0    &-  		&6650&   6686.4$\pm$31.0 &-36.4/-18.7& 10.3 &  Young Tail&\\
T124\tablenotemark{f}&  0    &-  		&6650&   6686.4$\pm$31.0 &-36.4/-18.7& 3.3 &  Young Tail&\\

\enddata
\tablenotetext{a}{\,0=Absorption , 1=emission}
\tablenotetext{b}{\,Lisenfeld et al. (2004)}
\tablenotetext{c}{\,Williams et al. (2002)}
\tablenotetext{d}{\,Difference between the cluster velocity measured here and the H{\sc i} velocity}
\tablenotetext{e}{\, From F11, Northern Starburst Region (NSBR), NGC~7318A/B, Southern Debris Region (SDR), Young Tail.}
\tablenotetext{e}{\,The properties of the TDG candidate, derived within an aperture of 35~pixels. 
The age is measured for the part of the clump covered in our spectroscopic slit, not the entire complex.}
\tablenotetext{f}{\,The properties of the TDG, this time confined to the region covered by our spectroscopic slit, as shown in Fig.~\ref{fig:tdg}.}

\label{table:properties2}
\end{deluxetable}

\begin{deluxetable}{lccc}
\tablecaption{Targets in the present work which were not confirmed to be members of the Stephan's Quintet system.  The first two columns give the ID in the present work and G01, respectively.  The third column gives the type of object, while the fourth column gives the category assigned to the target in G01.  'S' denotes sources classified as likely foreground stars in G01. }
\tablewidth{-10pt}

\tablehead{
\colhead{ID$_{T10A}$} & \colhead{ID$_{G01}$}  & \colhead{C$_{T10A}$}  & \colhead{C$_{G01}$\tablenotemark{a}}} 

\startdata
T480&54  &Quasar & B  \\
T654&64  &Star &D  \\
T234&42  &Star&D    \\
T288&50  &Star &D     \\
T2154&17&Star&S,T   \\
T2155&18&Star&S,T    \\
T2156&16&Star &S,T   \\
T2153&6 &Star &S   \\
T1195&129&Star  &S     \\
T1150&117&Star  &S     \\
T756&71    &Star&D     \\
T702&66    &Star &D  \\
\enddata
\tablenotetext{a}{\,The designation of the source in G01.  `T' means that the source is associated with the Young Tail.  The other designations relate to color groups, B: $0.1 \le B-V < 0.3$; D: $0.8 \le < B-V < 1.1$; and S (likely foreground stars): $1.1 \le B-V$.}

\label{table:comparison1}
\end{deluxetable}

\begin{deluxetable}{ll}

\tablecaption{Confirmed cluster candidates from the Galagher et al.~(2001) study. IDs follow Table~\ref{table:comparison2}.}
\tablewidth{-10pt}

\tablehead{
\colhead{ID$_{T10A}$} & \colhead{ID$_{G01}$}
}

\startdata
T110a&131    \\
T110b&123      \\
T110e&104       \\
T112&137       \\
T117& 7    \\
T118& 12      \\
T119& 76      \\
T120& 78      \\
T123&91     \\
\enddata

\label{table:comparison2}
\end{deluxetable}

\begin{deluxetable}{lccccccc}
\def\psn{\phs\phn}
\def\pnn{\phn\phn}
\tablecolumns{8}
\tablewidth{0pt}
\tablecaption{Measured Absorption Line Indices}
\tablehead{
\colhead{ID} &
\colhead{ H+He\tablenotemark{a}} &
\colhead{ K\tablenotemark{a}} &
\colhead{ H8\tablenotemark{a}} &
\colhead{ H$\gamma_A$\tablenotemark{b}}  &
\colhead{ Mgb5177\tablenotemark{b}} &
\colhead{ Fe5270\tablenotemark{b}}  &
\colhead{ Fe5335\tablenotemark{b}} \\
\colhead{} & \colhead{ (\AA) } & \colhead{ (\AA) } & \colhead{ (\AA) }  & \colhead{( \AA)}& \colhead{( \AA)}& \colhead{( \AA)}& \colhead{( \AA)} 
}
\startdata
T113&7.77$\pm$0.31&0.72$\pm$0.91&6.48$\pm$0.95&5.90$\pm$0.57&0.36$\pm$0.11&0.85$\pm$0.37&1.00$\pm$0.29\\      
T118&10.5$\pm$0.51&1.82$\pm$0.33&8.50$\pm$0.55&8.11$\pm$0.33&0.60$\pm$0.16&0.90$\pm$0.20&0.96$\pm$0.29\\       
T119&6.36$\pm$0.72&1.02$\pm$0.43&4.77$\pm$0.74&4.84$\pm$0.46&0.30$\pm$0.21&0.76$\pm$0.26&1.04$\pm$0.38\\       
T120&5.59$\pm$0.56&0.66$\pm$0.33&4.18$\pm$0.56&4.54$\pm$0.36&0.52$\pm$0.18&0.71$\pm$0.22&0.97$\pm$0.32\\       
T121&7.99$\pm$1.03&0.88$\pm$0.60&6.37$\pm$0.93&6.47$\pm$0.44&0.29$\pm$0.22&0.97$\pm$0.25&1.06$\pm$0.32\\       
T122&8.06$\pm$1.75&0.84$\pm$0.87&6.34$\pm$0.54&6.60$\pm$0.47&0.30$\pm$0.60&0.95$\pm$0.05&1.07$\pm$0.20\\       
T123&6.15$\pm$1.17&2.61$\pm$0.66&4.21$\pm$1.20&4.39$\pm$0.67&0.65$\pm$0.31&0.99$\pm$0.38&1.06$\pm$0.53\\       
\enddata
\tablenotetext{a}{\,Indices as defined by Schweizer \& Seitzer (1998).}
\tablenotetext{b}{\,Lick indices.}
\label{table:indices}
\end{deluxetable}

\begin{deluxetable}{lccccc}
\def\psn{\phs\phn}
\def\pnn{\phn\phn}
\tablecolumns{6}
\tablewidth{0pt}
\tablecaption{Measured Emission Line Lick Indices (in \AA)}
\tablehead{
\colhead{ID} &
\colhead{[OII]{$\lambda3727$}} &
\colhead{H$\gamma_A$} &
\colhead{H$\beta$}  &
\colhead{[OIII]$\lambda4959$}    &
\colhead{[OIII]$\lambda5007$}   
}
\startdata 
T110a      & -                 		&-23.79$\pm$1.09& -79.38$\pm$1.83&-37.47$\pm$0.44&-109.31$\pm$0.97\\  
T110b      & -                 		& -8.32$\pm$1.11&-64.39$\pm$0.98&-31.70$\pm$0.53&-90.72$\pm$0.99\\     
T110c      & -                 		&-15.67$\pm$9.83&-21.23$\pm$4.54&-11.54$\pm$3.31&-67.77$\pm$5.90\\  
T110d      & -                 		& -5.89$\pm$0.45&-22.04$\pm$0.36&-11.85$\pm$0.16&-44.44$\pm$0.28\\     
T110e      & -                 		&-18.42$\pm$0.40 &-85.09$\pm$0.23&-65.88$\pm$0.29&-226.63$\pm$0.68\\   
T111a      & -48.02$\pm$3.15&-14.02$\pm$0.93 & -58.79$\pm$1.44 &-19.78$\pm$0.51&-55.29$\pm$0.65\\   
T111b      &-47.03$\pm$7.70&-16.64$\pm$1.96 & -40.50$\pm$4.15 &-16.06$\pm$4.45&-43.74$\pm$5.29\\    
T111c      &-42.58$\pm$31.48& -1.21$\pm$4.23 & -22.29$\pm$5.30 &-13.75$\pm$3.79&-42.21$\pm$4.48\\      
T112        &-110.03$\pm$2.12&-14.24$\pm$0.32&-53.32$\pm$0.44&-35.21$\pm$0.20&-106.13$\pm$0.35\\    
T114        &-110.81$\pm$7.64&-14.39$\pm$0.88&-69.83$\pm$1.42&-57.02$\pm$0.66&-173.25$\pm$1.29\\ 
T115        & -26.20$\pm$1.40&-5.22$\pm$0.04&-12.01$\pm$0.23 &-3.61$\pm$0.05&-7.63$\pm$0.07\\             
T116b      & -23.67$\pm$0.11&-4.80$\pm$0.18&-18.27$\pm$0.04 &-11.00$\pm$0.05&-32.86$\pm$0.01\\           
T117        & -66.87$\pm$5.74&-31.18$\pm$0.84&-168.26$\pm$1.82 &-119.62$\pm$0.19&-273.02$\pm$0.50\\    

\enddata
\label{table:indices2}
\end{deluxetable}

\clearpage


\begin{thebibliography}

\bibitem[{{Appleton} {et~al.}(2006){Appleton}, {Xu}, {Reach}, {Dopita}, {Gao}, {Lu}, {Popescu}, {Sulentic}, {Tuffs}, \& {Yun}}]{Appleton2006} {Appleton}, P.~N., {Xu}, K.~C., {Reach}, W., {Dopita}, M.~A., {Gao}, Y., {Lu}, N., {Popescu}, C.~C., {Sulentic}, J.~W., {Tuffs}, R.~J., \& {Yun}, M.~S. 2006, \apjl, 639, L51

\bibitem[Baldwin et al.(1981)]{bpt} Baldwin, J.~A., 
Phillips, M.~M., \& Terlevich, R.\ 1981, \pasp, 93, 5 

\bibitem[Barmby et al.(2006)]{2006AJ....132..883B} Barmby, P., Kuntz, K.~D., Huchra, J.~P., \& Brodie, J.~P.\ 2006, \aj, 132, 883 

\bibitem[Bastian et al.(2005)]{2005A&A...435...65B} Bastian, N., Hempel, M., Kissler-Patig, M., Homeier, N.~L., \& Trancho, G.\ 2005, \aap, 435, 65 

\bibitem[Bastian(2008)]{2008MNRAS.390..759B} Bastian, N.\ 2008, \mnras, 390, 759 

\bibitem[Bastian et al.(2009)]{2009ApJ...701..607B} Bastian, N., Trancho, G., Konstantopoulos, I.~S., \& Miller, B.~W.\ 2009, \apj, 701, 607 (B09)

\bibitem[Baumgardt \& Makino(2003)]{2003MNRAS.340..227B} Baumgardt, H., \& Makino, J.\ 2003, \mnras, 340, 227 

\bibitem[Cardiel et al.~(1998)]{cardiel98} Cardiel, N., Gorgas, J.,
   Cenarro, J., Gonzalez, J.J. 1998, A\&AS, 127, 597
   
\bibitem[Cappellari \& Emsellem(2004)]{2004PASP..116..138C} Cappellari, M., 
\& Emsellem, E.\ 2004, \pasp, 116, 138 

\bibitem[Fedotov et al.(2011)]{2011arXiv1105.5840F} Fedotov, K., Gallagher, S.~C., Konstantopoulos, I.~S., Chandar, R., Bastian, N., Charlton, J.~C., Whitmore, B., \& Trancho, G.\ 2011, AJ, in press (arXiv:1105.5840) (F11)

\bibitem[{{Gallagher} {et~al.}(2001){Gallagher}, {Charlton}, {Hunsberger}, {Zaritsky}, \& {Whitmore}}]{Gallagher2001} {Gallagher}, S.~C., {Charlton}, J.~C., {Hunsberger}, S.~D., {Zaritsky}, D., \& {Whitmore}, B.~C. 2001, \aj, 122, 163

\bibitem[Gallagher et al.(2010)]{2010AJ....139..545G} Gallagher, S.~C., et al.\ 2010, \aj, 139, 545 (G01)

\bibitem[Gieles et al.(2006)]{2006A&A...450..129G} Gieles, M., Larsen, S.~S., Bastian, N., \& Stein, I.~T.\ 2006, \aap, 450, 129 

\bibitem[Gieles(2010)]{2010ASPC..423..123G} Gieles, M.\ 2010, in ASP conference series Vol 423, Galaxy Wars: Stellar Populations and Star Formation in Interacting Galaxies, 423, 123

\bibitem[Gonz{\'a}lez Delgado et al.(2005)]{2005MNRAS.357..945G} Gonz{\'a}lez Delgado, R.~M., Cervi{\~n}o, M., Martins, L.~P., Leitherer, C., \& Hauschildt, P.~H.\ 2005, \mnras, 357, 945 

\bibitem[{{Guillard} {et~al.}(2010){Guillard}, {Boulanger}, {Cluver}, {Appleton}, {Pineau Des For{\^e}ts}, \& {Ogle}}]{Guillard2010} {Guillard}, P., {Boulanger}, F., {Cluver}, M.~E., {Appleton}, P.~N., {Pineau Des For{\^e}ts}, G., \& {Ogle}, P. 2010, \aap, 518, A59+



\bibitem[Karl et al.(2010)]{2010ApJ...715L..88K} Karl, S.~J., Naab, T., Johansson, P.~H., Kotarba, H., Boily, C.~M., Renaud, F., \& Theis, C.\ 2010, \apjl, 715, L88 

\bibitem[Karl et al.(2011)]{2011ApJ...734...11K} Karl, S.~J., Fall, S.~M., \& Naab, T.\ 2011, \apj, 734, 11 

\bibitem[Kauffmann et al.(2003)]{2003MNRAS.346.1055K} Kauffmann, G., et al.\ 2003, \mnras, 346, 1055 

\bibitem[Kewley et al.(2006)]{2006MNRAS.372..961K} Kewley, L.~J., Groves, B., Kauffmann, G., \& Heckman, T.\ 2006, \mnras, 372, 961 

\bibitem[Kobulnicky \& Kewley(2004)]{2004ApJ...617..240K} Kobulnicky, H.~A., \& Kewley, L.~J.\ 2004, \apj, 617, 240 

\bibitem[Konstantopoulos et al.(2010)]{2010ApJ...723..197K} Konstantopoulos, I.~S., et al.\ 2010, \apj, 723, 197 

\bibitem[Kroupa(1998)]{1998ASPC..134..483K} Kroupa, P.\ 1998, Brown Dwarfs and Extrasolar Planets, 134, 483  

\bibitem[Kruijssen et al.(2011)]{2011MNRAS.414.1339K} Kruijssen, J.~M.~D., Pelupessy, F.~I., Lamers, H.~J.~G.~L.~M., Portegies Zwart, S.~F., \& Icke, V.\ 2011, \mnras, 414, 1339 

\bibitem[Lamers et al.(2005)]{2005A&A...441..117L} Lamers, H.~J.~G.~L.~M., Gieles, M., Bastian, N., Baumgardt, H., Kharchenko, N.~V., \& Portegies Zwart, S.\ 2005, \aap, 441, 117 

\bibitem[Larsen(1999)]{1999A&AS..139..393L} Larsen, S.~S.\ 1999, \aaps, 139, 393 

\bibitem[Larsen(2002)]{2002AJ....124.1393L} Larsen, S.~S.\ 2002, \aj, 124, 139

\bibitem[Larsen(2004)]{2004A&A...416..537L} Larsen, S.~S.\ 2004, \aap, 416, 537 

\bibitem[Larsen(2009)]{2009A&A...494..539L} Larsen, S.~S.\ 2009, \aap, 494, 539 

\bibitem[Leitherer et al.(1999)]{1999ApJS..123....3L} Leitherer, C., et al.\ 1999, \apjs, 123, 3 

\bibitem[{{Lisenfeld} {et~al.}(2004){Lisenfeld}, {Braine}, {Duc}, {Brinks}, {Charmandaris}, \& {Leon}}]{Lisenfeld2004} {Lisenfeld}, U., {Braine}, J., {Duc}, P.-A., {Brinks}, E., {Charmandaris}, V.,   \& {Leon}, S. 2004, \aap, 426, 471

\bibitem[Maraston et al.(2004)]{2004A&A...416..467M} Maraston, C., Bastian, N., Saglia, R.~P., Kissler-Patig, M., Schweizer, F., \& Goudfrooij, P.\ 2004, \aap, 416, 467 

\bibitem[Marigo et al.(2008)]{2008A&A...482..883M} Marigo, P., Girardi, L., Bressan, A., Groenewegen, M.~A.~T., Silva, L., \& Granato, G.~L.\ 2008, \aap, 482, 883 

\bibitem[{{Mendes de Oliveira} {et~al.}(2001){Mendes de Oliveira}, {Plana}, {Amram}, {Balkowski}, \& {Bolte}}]{Mendes2001} {Mendes de Oliveira}, C., {Plana}, H., {Amram}, P., {Balkowski}, C., \& {Bolte}, M. 2001, \aj, 121, 2524


\bibitem[Mieske et al.(2008)]{2008A&A...487..921M} Mieske, S., et al.\ 2008, \aap, 487, 921 

\bibitem[Mihos et al.(1993)]{1993ApJ...418...82M} Mihos, J.~C., Bothun, G.~D., \& Richstone, D.~O.\ 1993, \apj, 418, 82 


\bibitem[{{Moles} {et~al.}(1997){Moles}, {Sulentic}, \& {Marquez}}]{Moles1997} {Moles}, M., {Sulentic}, J.~W., \& {Marquez}, I. 1997, \apjl, 485, L69+

\bibitem[Natale et al.(2010)]{2010ApJ...725..955N} Natale, G., et al.\ 2010, \apj, 725, 955 

\bibitem[Plana et al.(1999)]{1999ApJ...516L..69P} Plana, H., Mendes de Oliveira, C., Amram, P., Bolte, M., Balkowski, C., \& Boulesteix, J.\ 1999, \apjl, 516, L69 

\bibitem[Portegies Zwart et al.(2010)]{2010ARA&A..48..431P} Portegies Zwart, S.~F., McMillan, S.~L.~W., \& Gieles, M.\ 2010, \araa, 48, 431 

\bibitem[Renaud et al.(2010)]{2010ApJ...724...80R} Renaud, F., Appleton, P.~N., \& Xu, C.~K.\ 2010, \apj, 724, 80 

\bibitem[Rupke et al.(2010)]{2010ApJ...723.1255R} Rupke, D.~S.~N., Kewley, L.~J., \& Chien, L.-H.\ 2010, \apj, 723, 1255 

\bibitem[Scheepmaker et al.(2007)]{2007A&A...469..925S} Scheepmaker, R.~A., Haas, M.~R., Gieles, M., Bastian, N., Larsen, S.~S., \& Lamers, H.~J.~G.~L.~M.\ 2007, \aap, 469, 925 

\bibitem[Schlegel et al.(1998)]{1998ApJ...500..525S} Schlegel, D.~J., Finkbeiner, D.~P., \& Davis, M.\ 1998, \apj, 500, 525 

\bibitem[Schweizer \& Seitzer(1998)]{1998AJ....116.2206S} Schweizer, F., \& Seitzer, P.\ 1998, \aj, 116, 2206 

\bibitem[Schweizer et al.(2004)]{2004AJ....128..202S} Schweizer, F., Seitzer, P., \& Brodie, J.~P.\ 2004, \aj, 128, 202 

\bibitem[{{Sulentic} {et~al.}(2001){Sulentic}, {Rosado}, {Dultzin-Hacyan}, {Verdes-Montenegro}, {Trinchieri}, {Xu}, \& {Pietsch}}]{Sulentic2001} {Sulentic}, J.~W., {Rosado}, M., {Dultzin-Hacyan}, D., {Verdes-Montenegro}, L., {Trinchieri}, G., {Xu}, C., \& {Pietsch}, W. 2001, \aj, 122, 2993

\bibitem[{{Suzuki} {et~al.}(2011){Suzuki}, {Kaneda}, {Onaka}, \&  {Kitayama}}]{Suzuki2011} {Suzuki}, T., {Kaneda}, H., {Onaka}, T., \& {Kitayama}, T. 2011, \apjl, 731,  L12+

\bibitem[{{Torres-Flores} {et~al.}(2009){Torres-Flores}, {Mendes de Oliveira},  {de Mello}, {Amram}, {Plana}, {Epinat}, \&  {Iglesias-P{\'a}ramo}}]{Torres2009} {Torres-Flores}, S., {Mendes de Oliveira}, C., {de Mello}, D.~F., {Amram}, P., {Plana}, H., {Epinat}, B., \& {Iglesias-P{\'a}ramo}, J. 2009, \aap, 507, 723

\bibitem[Tran et al.(2003)]{2003ApJ...585..750T} Tran, H.~D., et al.\ 2003, \apj, 585, 750 

\bibitem[Trancho et al.(2007)]{2007ApJ...658..993T} Trancho, G., Bastian, N., Schweizer, F., \& Miller, B.~W.\ 2007a, \apj, 658, 993 (T07a)

\bibitem[Trancho et al.(2007)]{2007ApJ...664..284T} Trancho, G., Bastian, N., Miller, B.~W., \& Schweizer, F.\ 2007b, \apj, 664, 284 (T07b)

\bibitem[{{Trinchieri} {et~al.}(2003){Trinchieri}, {Sulentic}, {Breitschwerdt},   \& {Pietsch}}]{Trinchieri2003} {Trinchieri}, G., {Sulentic}, J., {Breitschwerdt}, D., \& {Pietsch}, W. 2003, \aap, 401, 173


\bibitem[Werk et al.(2008)]{2008ApJ...678..888W} Werk, J.~K., Putman, M.~E., Meurer, G.~R., Oey, M.~S., Ryan-Weber, E.~V., Kennicutt, R.~C., Jr., \& Freeman, K.~C.\ 2008, \apj, 678, 888 

\bibitem[West et al.(1995)]{1995ApJ...453L..77W} West, M.~J., Cote, P., Jones, C., Forman, W., \& Marzke, R.~O.\ 1995, \apjl, 453, L77 

\bibitem[Whitmore(2003)]{whitmore03} Whitmore, B. C. 2003, in A Decade of Hubble Space Telescope Science, 153-178

\bibitem[{{Williams} {et~al.}(2002){Williams}, {Yun}, \& {Verdes-Montenegro}}]{Williams2002} {Williams}, B.~A., {Yun}, M.~S., \& {Verdes-Montenegro}, L. 2002, \aj, 123, 2417

\bibitem[{{Xu} {et~al.}(1999){Xu}, {Sulentic}, \& {Tuffs}}]{Xu1999} {Xu}, C., {Sulentic}, J.~W., \& {Tuffs}, R. 1999, \apj, 512, 178

\bibitem[{{Xu} {et~al.}(2005){Xu}, {Iglesias-P{\'a}ramo}, {Burgarella}, {Rich}, {Neff}, {Lauger}, {Barlow}, {Bianchi}, {Byun}, {Forster}, {Friedman}, {Heckman}, {Jelinsky}, {Lee}, {Madore}, {Malina}, {Martin}, {Milliard}, {Morrissey}, {Schiminovich}, {Siegmund}, {Small}, {Szalay}, {Welsh}, \& {Wyder}}]{Xu2005} {Xu}, C.~K., {Iglesias-P{\'a}ramo}, J., {Burgarella}, D., {Rich}, R.~M., {Neff}, S.~G., {Lauger}, S., {Barlow}, T.~A., {Bianchi}, L., {Byun}, Y.-I.,   {Forster}, K., {Friedman}, P.~G., {Heckman}, T.~M., {Jelinsky}, P.~N., {Lee},   Y.-W., {Madore}, B.~F., {Malina}, R.~F., {Martin}, D.~C., {Milliard}, B.,   {Morrissey}, P.,  {Schiminovich}, D., {Siegmund}, O.~H.~W., {Small}, T.,   {Szalay}, A.~S., {Welsh}, B.~Y., \& {Wyder}, T.~K. 2005, \apjl, 619, L95

\bibitem[{{Xu} {et~al.}(2003){Xu}, {Lu}, {Condon}, {Dopita}, \&   {Tuffs}}]{Xu2003}
{Xu}, C.~K., {Lu}, N., {Condon}, J.~J., {Dopita}, M., \& {Tuffs}, R.~J. 2003, \apj, 595, 665


\end{thebibliography}
\end{document}